\documentclass[review]{elsarticle}

\usepackage{%lineno,
hyperref,amsmath, amssymb}
\usepackage[strict]{changepage}
\usepackage{xcolor}
\renewcommand{\(}{\begin{equation}}
\renewcommand{\)}{\end{equation}}
\newcommand{\bea}{\begin{eqnarray}}
\newcommand{\eea}{\end{eqnarray}}

\newcommand{\beq}{\begin{equation}}
\newcommand{\eeq}{\end{equation}}
\renewcommand{\(}{\begin{equation}}
\renewcommand{\)}{\end{equation}}

\def\bp{\mbox{\boldmath $p$}}
\def\bSigma{\mbox{\boldmath $\Sigma$}}

\newcommand{\Mov}[1]{{%\bf
\color{black}{#1}}}

\begin{document}

\begin{frontmatter}

%\title{Correlations in CLASH Clusters}%Elsevier \LaTeX\ template\tnoteref{mytitlenote}}
\title{Correlations in the matter distribution in CLASH galaxy clusters}%Elsevier \LaTeX\ template\tnoteref{mytitlenote}}
%\tnotetext[mytitlenote]{Fully documented templates are available in the elsarticle package on \href{http://www.ctan.org/tex-archive/macros/latex/conadtrib/elsarticle}{CTAN}.}
%é,
%% Group a&,aduthors per affiliation:
\author[ad1,ad2]{Antonino Del Popolo}%Elsevier\fnref{myfootnote}}
\ead{adelpopolo@oact.inaf.it}
\address[ad1]{Dipartimento di Fisica e Astronomia, University Of Catania, Viale Andrea Doria 6, 95125 Catania, Italy}
\address[ad2]{INFN sezione di Catania, Via S. Sofia 64, I-95123 Catania, Italy}
%\fntext[myfootnote]{Since 1880.}

%% or include affiliations in footnotes:
\author[adm1,adm2]{Morgan Le Delliou\corref{corauth}}
%\ead[url]{www.elsevier.com}
\cortext[corauth]{Corresponding author}
\ead{delliou@lzu.edu.cn,Morgan.LeDelliou.IFT@gmail.com}
  
%\author[mysecondaryaddress]{Global Customer Service\corref{mycorrespondingauthor}}
%\cortext[mycorrespondingauthor]{Corresponding author}
%\ead{support@elsevier.com}

\address[adm1]{Institute of Theoretical Physics, School of Physical Science and Technology, Lanzhou University, No.222, South Tianshui Road, Lanzhou, Gansu, 730000, Peoples Republic of China}
\address[adm2]{Instituto de Astrof\'ısica e Ci\^encias do Espa\c co, Universidade de Lisboa, Faculdade de Ciˆencias, Ed. C8, Campo Grande, 1769-016 Lisboa, Portugal}

\author[ad3]{Xiguo Lee\fnref{fn1}}\fntext[fn1]{This paper is dedicated to the memory of our colleague Prof. Lee, Xiguo}

\address[ad3]{Institute of Modern Physics, Chinese Academy of Sciences, Post Office Box31, Lanzhou 730000, Peoples Republic of China}

\begin{abstract}
We study the total and dark matter (DM) density profiles as well as
their correlations for a sample of 15 high-mass galaxy clusters by
extending our previous work on several clusters from Newman et al. 
Our analysis focuses on 15 CLASH X-ray-selected clusters that have
high-quality weak- and strong-lensing measurements from combined Subaru
and {\em Hubble Space Telescope} observations. 
The total density profiles derived from lensing are interpreted based on
the two-phase scenario of cluster formation.
In this context, the brightest cluster galaxy (BCG) forms in the first
dissipative phase, followed by a dissipationless phase where
baryonic physics flattens the inner DM distribution. 
This results in the formation of clusters
with modified DM distribution and several correlations between 
characteristic quantities of the clusters.    
We find that the central DM density profiles of the clusters are
strongly influenced by baryonic physics as found in our earlier work.
The inner slope of the DM density for the CLASH clusters is found to  
be flatter than the Navarro--Frenk--White profile,  ranging
from $\alpha=0.30$ to $0.79$.
We examine correlations of the DM density slope $\alpha$ with the effective radius $R_\mathrm{e}$ and
stellar mass $M_\mathrm{e}$ of the BCG, finding that
these quantities are anti-correlated with a Spearman correlation
coefficient of $\sim -0.6$. 
We also study the correlation between $R_\mathrm{e}$ and the
cluster halo mass $M_{500}$, and the correlation between the
total masses inside 5\,kpc and 100\,kpc. We find that
these quantities are correlated with
Spearman coefficients of $0.68$ and $0.64$, respectively.
These observed correlations are in support of the physical picture
proposed by Newman et al.
\end{abstract}

\begin{keyword}
  Galaxy Clusters, Galaxy Formation, Weak Gravitational Lensing
  %\texttt{elsarticle.cls}\sep \LaTeX\sep Elsevier \sep template
%\MSC[2010] 00-01\sep  99-00
\end{keyword}

\end{frontmatter}

%\linenumbers

\section{Introduction}
The $\Lambda$ cold dark matter ($\Lambda$CDM) paradigm gives a plethora
of correct predictions
\citep{Komatsu2011,Planck2014,DelPopolo2013,DelPopolo2007,DelPopolo2014aa}. However,
some of its predictions are at odds with observations.
$N$-body simulations in $\Lambda$CDM predict that the spherically
averaged density
profiles of self-gravitating structures, ranging from dwarf galaxies
to galaxy clusters, are cuspy and well approximated by the
Navarro--Frenk--White (NFW) profile \citep{Navarro1997,Navarro2010}.
However,
observations \citep{Moore1994,Flores1994,Agnello2012,Adams2014} and theoretical studies
\citep{Navarro1996,DelPopolo2009,Governato2010,DelPopolo2010,DelPopolo2011,DelPopolo2012a}
%({\bf KU: Here I added more references}
have shown that the inner slopes of the density profile in dwarf
galaxies and low-surface-brightness galaxies (LSBs) are usually flatter
than simulations, and there is a strong diversity of the dark-matter
(DM) distribution in these low-mass systems \citep[the so-called ``diversity
problem'',][]{Simon2005,DelPopolo2012a,Oman2015}.\footnote{In addition to this problem,
the $\Lambda$CDM paradigm suffers from the cosmological constant problem
\citep{Weinberg1989,Astashenok2012}, the unknown nature of dark energy
\citep{DelPopolo2013a,DelPopolo2013b,DelPopolo2013c}, and from several
problems at small scales \citep{DelPopolo2017,DelPopolo2017a}}

On the observational side,
the small dynamic range of observations can cause a degeneracy in
the mass profile determination \citep[see][]{DelPopolo2002},
and this degeneracy cannot be fully broken due to the lack of HI observations in dwarf
spheroidals (dSPhs) and elliptical galaxies. Determinations of their DM
structure are thus much more complicated. In the case of dSPhs, 
there are discrepant results on the cusp-core nature of the density profile   
\citep{Amorisco2012,Jardel2012,Jardel2013a,Jardel2013b}, sometimes even
in the case of the same object studied with different techniques. 
Similar uncertainties are present in cluster of galaxies, but X-ray
observations, lensing and galaxies dynamics overcome them in an easier
manner than for the cases of dSPhs or ellipticals.  

While the NFW profile
\citep{Navarro1996,Navarro1997} describes well the observed total
density profiles in galaxy clusters as found in several studies
\citep{Sand2002,Sand2004,Newman2009,Umetsu2011,Newman2011,Okabe2013,Newman2013a,Newman2013b,Umetsu2014,Umetsu2016},
it was also found that the inner DM structure is characterized by a
flatter slope within typical scales of the brightest cluster
galaxy (BCG; from some kpcs to some tens of kpcs).
Hence, the cusp-core problem \citep{DelPopolo2012a,Oman2015} appears to
be present in galaxy clusters as well.
This discrepancy can be alleviated when the effects of baryonic physics
are properly accounted for in $N$-body simulations
\citep[see][]{ElZant2004,Nipoti2004,DelPopolo2009,DelPopolo2009a,DelPopolo2010,Cardone2011a,Cardone2011b,Governato2010,Cole2011,
DelPopolo2011,Martizzi2012,Martizzi2013,Nipoti2015}.

% Table generated by Excel2LaTeX from sheet 'Foglio1'
\begin{table}[htbp]
  \centering
  \caption{Parameters derived for the CLASH sample.
 First column: cluster name;
 second: $M_{500}$ as given in \citep{Umetsu2016};
 third and fourth: innermost 2D density slopes inferred 
 directly from the observed \citep{Umetsu2016} profiles and
 obtained from our semi-analytical model;
 fifth: inner 3D density slope from our model;
 sixth and seventh:
 stellar and baryonic fractions from our model. 
}
\small
    \begin{tabular}{lllllll}
\hline    
Name  & $M_{500}$ & $\alpha_\mathrm{2D}$ & $\alpha_\mathrm{2D,T}$ &
     $\alpha_\mathrm{3D}$ & $f_\mathrm{star}$  & $F_\mathrm{b}$ \\
    & [$10^{14}M_\odot$] & & & & & \\
\hline   
    A383  & 5.88  $\pm$1.73 & 0.71  $\pm$0.26 & 0.70  $\pm$0.09 & 0.37  $\pm$0.09 & 0.0201  $\pm$0.002 & 0.1355  $\pm$0.008 \\
    A209  & 9.64  $\pm$1.97 & 0.67  $\pm$0.29 & 0.68  $\pm$0.09 & 0.60  $\pm$0.1 & 0.0167  $\pm$0.002 & 0.1417  $\pm$0.01 \\
    A2261 & 15.65  $\pm$3.05 & 0.77  $\pm$0.26 & 0.79  $\pm$0.09 & 0.63  $\pm$0.09 & 0.0140  $\pm$0.002 & 0.1480  $\pm$0.012 \\
    RXJ2129 & 4.48  $\pm$1.16 & 0.49  $\pm$0.26 & 0.49  $\pm$0.09 & 0.55  $\pm$0.09 & 0.0222  $\pm$0.002 & 0.1323  $\pm$0.006 \\
    A611  & 10.73  $\pm$2.65 & 0.59  $\pm$0.27 & 0.58  $\pm$0.09 & 0.79  $\pm$0.09 & 0.0161  $\pm$0.002 & 0.1431  $\pm$0.01 \\
    MS2137 & 8.28  $\pm$2.57 & 0.86  $\pm$0.25 & 0.85  $\pm$0.09 & 0.65  $\pm$0.08 & 0.0177  $\pm$0.002 & 0.1398  $\pm$0.009 \\
    RXJ2248 & 12.45  $\pm$3.62 & 0.45  $\pm$0.28 & 0.44  $\pm$0.09 & 0.55  $\pm$0.09 & 0.0152  $\pm$0.002 & 0.1450  $\pm$0.011 \\
    MACSJ1115 & 10.67  $\pm$2.22 & 0.33  $\pm$0.30 & 0.34  $\pm$0.09 & 0.39  $\pm$0.1 & 0.0161  $\pm$0.002 & 0.1430  $\pm$0.01 \\
    MACSJ1931 & 10.51  $\pm$4.05 & 0.69  $\pm$0.28 & 0.70  $\pm$0.09 & 0.65  $\pm$0.09 & 0.0162  $\pm$0.002 & 0.1428  $\pm$0.01 \\
    MACSJ1720 & 9.96  $\pm$2.53 & 0.59  $\pm$0.26 & 0.60  $\pm$0.09 & 0.56  $\pm$0.09 & 0.0165  $\pm$0.002 & 0.1421  $\pm$0.01 \\
    MACSJ0429 & 6.85  $\pm$2.1 & 0.45  $\pm$0.28 & 0.44  $\pm$0.09 & 0.48  $\pm$0.09 & 0.0190  $\pm$0.002 & 0.1374  $\pm$0.008 \\
    MACSJ1206 & 12.24  $\pm$2.49 & 0.56  $\pm$0.26 & 0.57  $\pm$0.09 & 0.50  $\pm$0.09 & 0.0153  $\pm$0.002 & 0.1448  $\pm$0.011 \\
    MACSJ0329 & 6.51  $\pm$1.37 & 0.65  $\pm$0.27 & 0.64  $\pm$0.09 & 0.70  $\pm$0.09 & 0.0193  $\pm$0.002 & 0.1368  $\pm$0.008 \\
    RXJ1347 & 22.33  $\pm$4.89 & 0.39  $\pm$0.30 & 0.40  $\pm$0.09 & 0.30  $\pm$0.1 & 0.0123  $\pm$0.002 & 0.1528  $\pm$0.014 \\
    MACSJ0744 & 11.94  $\pm$2.81 & 0.54  $\pm$0.27 & 0.53  $\pm$0.09 & 0.55  $\pm$0.09 & 0.0155  $\pm$0.002 & 0.1445  $\pm$0.011 \\
    \hline
    \end{tabular}%
  \label{tab:1}%
\end{table}%

% Table generated by Excel2LaTeX from sheet 'Foglio1'
\begin{table}[htbp] 
  \centering
 \caption{Physical parameters derived for the CLASH sample.
 First column: cluster name;
 second: BCG mass derived from our model;
 third: BCG effective radius;
 fourth and fifth: spherical total masses inside 5\,kpc and 100\,kpc.}
\small
    \begin{tabular}{lllll} 
\hline
Name & $M_\mathrm{e}$ & $R_\mathrm{e}$ & $M_\mathrm{5 kpc}$ & $M_\mathrm{100 kpc}$\\
     & [$10^{11}M_\odot$] & [kpc] & [$10^{11}M_\odot$] & [$10^{13}M_\odot$]\\
    \hline    
    A383  & 9.16  $\pm$0.29 & 28.7  $\pm$1.5 & 0.98  $\pm$0.15 & 1.96  $\pm$0.3 \\
    A209  & 7.84  $\pm$0.29 & 25  $\pm$1.5 & 1.31  $\pm$0.15 & 3.21  $\pm$0.3 \\
    A2261 & 10.5  $\pm$0.29 & 40  $\pm$1.5 & 1.75  $\pm$0.15 & 4.32  $\pm$0.3 \\
    RXJ2129 & 13.73  $\pm$0.29 & 33  $\pm$1.5 & 2.43  $\pm$0.15 & 3.49  $\pm$0.3 \\
    A611  & 12.25  $\pm$0.29 & 34.6  $\pm$1.5 & 1.63  $\pm$0.15 & 3  $\pm$0.3 \\
    MS2137 & 6.55  $\pm$0.29 & 14  $\pm$1.5 & 1.95  $\pm$0.15 & 3.5  $\pm$0.3 \\
    RXJ2248 & 12.45  $\pm$0.29 & 38.5  $\pm$1.5 & 2.17  $\pm$0.15 & 4.3  $\pm$0.3 \\
    MACSJ1115 & 11.44  $\pm$0.29 & 44.5  $\pm$1.5 & 2  $\pm$0.15 & 3.56  $\pm$0.3 \\
    MACSJ1931 & 8.19  $\pm$0.29 & 31  $\pm$1.5 & 1.37  $\pm$0.15 & 3.5  $\pm$0.3 \\
    MACSJ1720 & 8.99  $\pm$0.29 & 35.8  $\pm$1.5 & 1.5  $\pm$0.15 & 3.32  $\pm$0.3 \\
    MACSJ0429 & 13.37  $\pm$0.29 & 41  $\pm$1.5 & 2.32  $\pm$0.15 & 3.28  $\pm$0.3 \\
    MACSJ1206 & 11.96  $\pm$0.29 & 43  $\pm$1.5 & 2.08  $\pm$0.15 & 4.08  $\pm$0.3 \\
    MACSJ0329 & 7.41  $\pm$0.29 & 20  $\pm$1.5 & 1.24  $\pm$0.15 & 2.17  $\pm$0.3 \\
    RXJ1347 & 13.8  $\pm$0.29 & 46.9  $\pm$1.5 & 2.45  $\pm$0.15 & 4.5  $\pm$0.3 \\
    MACSJ0744 & 10  $\pm$0.29 & 37.1  $\pm$1.5 & 1.67  $\pm$0.15 & 3.98  $\pm$0.3 \\
\hline    
    \end{tabular}%
  \label{tab:2}%
\end{table}%

In order to study how baryons modify the formation and evolution of
clusters, we consider in \citep{DelPopolo2012b} baryonic clumps
interacting with the DM model introduced in \citep{DelPopolo2009}.
In addition to finding that the central baryonic concentration within
10\,kpc plays an important role in shaping the cluster density profile,
%and in correlations of the inner profile from the baryon content,
we reproduced the observed cluster profiles for several massive systems
\citep{Sand2004,Newman2009,Newman2011}, 
namely A611, A383, MACSJ1423.8+2404, and RXJ1133.  

In \citep{DelPopolo2014}, we reproduced the correlations found by
\citep{Newman2013a,Newman2013b},\footnote{The quoted authors
found correlations of the inner slope of the DM profile with
the size of the BCG, the core radius, namely the constant density core
of the cored NFW density profile \citep[see Eq.~2 of][]{Newman2013b}, 
and the BCG mass,  and finally the correlation between the masses
contained inside 5\,kpc and 100\,kpc. {%\bf 
In the present paper, with BCG mass we refer to the stellar mass only.}}  
for MS2137, A963, A383, A611, A2537, A2667, and A2390. 
For these clusters, the total mass density profiles 
are well fitted by an NFW profile, while the central DM distribution is
shallower than the total mass distribution. 

The formation picture proposed by \citep{Newman2013a,Newman2013b} is
characterized by a dissipational formation of BCGs, followed by a
dissipationless phase. In this phase, as described by
\citep{ElZant2004,Ma2004,Nipoti2004,RomanoDiaz2008,RomanoDiaz2009,DelPopolo2009,Inoue2011, 
DelPopolo2012a,DelPopolo2012b,Cole2011,Nipoti2015},
%Laporte2012,
baryon clumps interact with DM through dynamical friction, ``heating''
DM and reducing the central cusp. 

Our aims here are to use high-quality gravitational lensing observations 
from the CLASH survey \citep{Postman2012}
and investigate if CLASH clusters
exhibit correlations that are similar to those 
observed in the \citep{Newman2013a,Newman2013b}
clusters, to characterize the mass distributions of CLASH clusters, and
to test the physical picture that was proposed by
\citep{Newman2013a,Newman2013b} and confirmed by \citep{DelPopolo2014}. 
To this end, we perform an improved analysis on a sample of 15
X-ray-selected CLASH clusters compared to our previous work
\citep{DelPopolo2012b,DelPopolo2014}.

In the present work, we will characterize the total mass density profiles
of 15 CLASH clusters
by means of a modified version of the semi-analytical model developed by 
\citep{DelPopolo2012b,DelPopolo2014}. Here we take into account the
following effects: 
\begin{enumerate}
\item adiabatic contraction (AC) responsible for the steepening of the
      inner density profiles in the early stage of cluster formation,
\item interaction between baryonic clumps and DM through dynamical
      friction, which is responsible for ``heating'' the DM component
      and flattening the density profile, 
\item supernovae (SN) feedback,
\item AGN feedback and other baryonic effects described in detail in
      Appendix.
\end{enumerate}

The paper is organized as follows. In section \ref{sec:DataModel}, we
describe the data used and provide a brief summary of our model. In
section \ref{sec:Results}, we discuss the results, and section 
\ref{sec:Conclusions} is devoted to conclusions. 

%({\bf KU: The adopted cosmology must be stated! I assume that you used the same cosmology as in U16})
Throughout this paper, we adopt a concordance $\Lambda$CDM cosmology
with $\Omega_\mathrm{m}=0.27$, $\Omega_\Lambda=0.73$, and
$h=0.7$ with $H_0=100h$\,km\,s$^{-1}$\,Mpc$^{-1}$.

\section{Data used and summary of the model}\label{sec:DataModel}

In this study, we use lensing data 
obtained from the CLASH survey \cite{Postman2012}, which studied the
mass distributions of 25 high-mass clusters using high-quality
gravitational lensing observations. 
Here we focus on a subsample of 15 X-ray-regular CLASH clusters for
which strong and weak lensing data are available from
both 16-band {\em Hubble Space Telescope} ({\em HST}) and wide-field 
weak-lensing observations \citep{Umetsu2016}.
The wide-field weak-lensing data were taken primarily with Suprime-Cam  
on the Subaru Telescope \citep{Umetsu2014,Merten2015,Umetsu2018}.

We exclude high-magnification CLASH clusters from our analysis because
they are found to be highly disturbed merging systems \cite{Postman2012}.
We also exclude one X-ray-regular CLASH cluster (RXJ1532) for
which no secure identification of multiple images has been
made \citep{Zitrin2015} and hence no central strong-lensing information
is available. These selection criteria result in our sample of 15 CLASH
clusters (Table \ref{tab:1}). 

The data we use are taken from \citep{Umetsu2016} and given
in the form of binned surface mass density profiles,
spanning the radial range from 10\,arcsec to 16\,arcmin, and 
their bin-to--bin covariance matrices.
It was found in \citep{Umetsu2016} that the ensemble-averaged surface
mass density profile of these 
X-ray-regular clusters can be well described by cuspy, sharply 
steepening density profiles, such as the NFW and Einasto profiles.
Assuming the spherical NFW profile for each cluster,
\citep{Umetsu2016} also found that the concentration--mass relation for
the CLASH X-ray-selected subsample is in agreement with $\Lambda$CDM
predictions, when the CLASH selection function is taken into account.

In \citep{Umetsu2016}, the binned surface mass density profiles were
derived for a sample of 16 X-ray-regular and 4 
high-magnification CLASH clusters using the weak- and strong-lensing
data of \citep{Umetsu2014,Zitrin2015}. The mass profile solution for
each cluster, $\bSigma=\{\Sigma_i\}_{i=1}^N$ with $N=15$ bins, was
obtained from a joint likelihood analysis of strong-lensing,
weak-lensing shear and magnification data \citep[][see their
Fig.~11]{Umetsu2016}.
The total covariance matrix
$C_{ij}$ accounts for the observational errors,
the cosmic-noise contribution due to projected
uncorrelated large scale structure, the systematic errors due to the
residual mass-sheet degeneracy, and the intrinsic variations of 
the projected cluster mass profile due to halo triaxiality and
correlated substructures.

In this paper, we generate 3D cluster density
profiles whose surface density profiles match those
obtained by \citep{Umetsu2016}, $\bSigma$.
For a given 3D density profile $\rho(r)$,
we compute the surface mass density $\Sigma(R)$
by integrating $\rho(r)$ along the line of sight,
\begin{equation}
\Sigma(R)=2 \int_0^\infty \rho(R,l) dl
\end{equation}
with $R$ the projected cluster-centric radius \citep[see also Secs.~5.2
and 5.2.2 of][]{Umetsu2016}.  

A set of the parameters $\bp$ that specify the model
can be inferred by minimizing the $\chi^2$ function
\citep[see][]{Umetsu2016}, 
\begin{equation}
\chi^2(\bp)=\sum_{i,j=1}^{N}[\Sigma_i-\hat{\Sigma}_i(\bp)]C_{ij}^{-1} [\Sigma_j-\hat{\Sigma}_j(\bp)],
\label{Chi}
\end{equation}
where $\hat{\Sigma}_i$ is the surface mass density predicted by the model.

Estimates of cluster mass and its radial distribution can be obtained in
different ways.
The standard approach, as adopted by \citep{Umetsu2016}, is to use the
NFW profile, which gives a good approximation to the 
projected total density profile 
for cluster-size halos out to their virial radius
\citep{Oguri2011}. 
The NFW density profile is given by
\begin{equation}
\rho(r)=\frac{\rho_\mathrm{c} \delta_\mathrm{c}}{(r/r_\mathrm{s})(1+r/r_\mathrm{s})^2},
\label{NFW}
\end{equation}
where $r_\mathrm{s}=r_\Delta/c_\Delta$ is the scale
radius, $c_\Delta$ the concentration parameter,
$\rho_\mathrm{c}$ the critical density of the universe,
$r_\Delta$ the radius inside which the mean density is
$\Delta\times \rho_\mathrm{c}$, and 
\begin{equation}
\delta_\mathrm{c}=\frac{\Delta}{3}\frac{c_\Delta^3}{\ln(1+c_\Delta)-c_\Delta/(1+c_\Delta)}.
\end{equation}
The total mass enclosed within a sphere of radius $r_\Delta$ is denoted
as $M_\Delta=(4\pi/3)\Delta\rho_\mathrm{c}r_\Delta^3$.
The NFW mass and concentration parameters for the CLASH sample are
reported in Tables 2 and 3 of \citep{Umetsu2016}. The typical mass and
concentration for the X-ray-selected CLASH sample are
$M_{200}\simeq1.0\times 10^{15}M_\odot\,h^{-1}$ and
$c_{200}\simeq 3.8$ at $z\sim 0.35$
\citep{Umetsu2016,Umetsu2017}.

 In our model, described in %Appendix 
\ref{sec:model}, we %did not include the clustering 2-halo term, 
chose not to include the 2-halo clustering term, that takes
%which is the contribution from the surrounding large-scale structure around the main cluster
%to take 
into account the contribution on the projected surface mass density coming from the large-scale clustering.
As seen in Figure 3 of \citep{Umetsu2016}, the NFW approximation to the surface mass density $\Sigma(R)$ %is a good 
gives a good estimate %approximation 
to the virial %radius 
or to the $r_{200m}$ radius. When the fitting is restricted to the radial range $\leq 2Mpc/h$, the contribution from the 2-halo term is not important (\citep{Umetsu2014,Umetsu2016}). In other words, 
%if your 
since the %our 
results remain the same when our fitting range is limited to $<2Mpc/h$, our results are not sensitive to this effect.

Although the matter distributions in CLASH clusters is not spherical but triaxial \citep[see][]{Umetsu2018,Chiu2018},
%Umetsu et al. 2018; Sereno et al. 2018, ApJL; Chiu et al. 2018), 
as expected for dark-matter dominated cluster-scale halos \citep[e.g.][]{Meneghetti2014}, we assumed spherical symmetry in the constraints we obtained. %The uncertainty arising from projection effects of aspherical cluster shapes, our spherical modeling of CLASH clusters is accounted for
%
%Moreover, we assumed spherical symmetry in the constraints we obtained. We know that the matter distributions in CLASH clusters are not spherical but triaxial (see \citep{Umetsu2018,Chiu2018}),
%Umetsu et al. 2018; Sereno et al. 2018, ApJL; Chiu et al. 2018), 
%as expected for dark-matter dominated cluster-scale halos (e.g., \citep{Meneghetti2014}). 
\Mov{In our spherical modeling of CLASH clusters, we account for the uncertainty arising from projection effects of aspherical cluster shapes 
in the covariance matrix $C^\mathrm{int}$ \citep[see][]{Umetsu2016}.}

In order to characterize the 3D density profile of the clusters, we use
a modified version of the physical cluster model described in \citep{DelPopolo2012b,DelPopolo2014}.
In those papers \citep[i.e.][]{DelPopolo2012b,DelPopolo2014}, 
%{\bf Here, we discuss the difference between the method used to constrain our model for each of the CLASH %clusters, to obtain the density profiles and correlations in \citep{DelPopolo2012b,DelPopolo2014} and that %of this paper.}
%We constrain our model for each of the CLASH clusters as follows.
%In \citep{DelPopolo2012b,DelPopolo2014}, 
the DM density  profile
was expressed as $\rho_\mathrm{DM}= F(M_{500},F_\mathrm{b},j)$,
with
$M_{500}$ being the cluster halo mass,
$f_\mathrm{b}=M_\mathrm{b}/M_{500}$ the cluster baryon fraction,
and $j$ the random angular momentum 
%({\bf KU: I explicitly show that your mass is M500 here. M500 is not virial mass}).  
%In \citep{DelPopolo2012b,DelPopolo2014}, 
The density profile of the
clusters studied were reproduced by
a) assuming that the cluster final mass in the model is the same as the
observed clusters,
b) assuming that  the cluster baryon fraction 
%that in \citep{DelPopolo2012b,DelPopolo2014} was obtained using 
is equal to that calculated with the \citep{Giodini2009} data,
%\footnote{The virial mass $M_{200}$ was converted to $M_{500}$ following...}, 
and c) adjusting the value of the random angular momentum to reproduce
the observed clusters profiles\footnote{We recall that clusters of
galaxies are not supported by rotation, and that their ``ordered'' angular
momentum, coming from tidal torques, has very similar and low values
\citep[some km\,s$^{-1}$][]{Catelan1996a,Catelan1996b}, for all clusters, and in
terms of the spin parameter $\lambda$ can be fixed to the typical value
$\lambda= 0.03$ \citep{Gottlober2007}.}. 

{%\bf 
The data in the present study does not provide the DM density profile, while we have the total density profile from lensing, and the profile of gas mass recovered from literature.
Being the last the major contributor to the baryonic component, it is needed, and used to constrain the baryon fraction.
}
%
% and baryon density profiles themselves but only the total density profile from lensing.
%
Hence, unlike \citep{DelPopolo2012b,DelPopolo2014},
the DM content for the CLASH clusters
cannot be extracted %determined 
from fitting the model to the total and baryon
density profiles separately.   
%In order to derive the DM density profiles for the CLASH clusters, we subtract from the total density (given %by the data) the baryon density given by the model.  

The %In the 
present paper reproduces %, 
the CLASH density profiles %have been reproduced 
in
a different manner. 
In the same way as %As 
in \citep{DelPopolo2012b,DelPopolo2014}, we assume that the final
mass of the protostructure generating each of the CLASH clusters is
equal to the observed mass of that cluster, namely $M_{500}$ from
\citep{Umetsu2016}, and that the surface density given from observation is reproduced by the model.
However and contrary to \citep{DelPopolo2012b,DelPopolo2014}, the %The 
cluster baryon fraction $F_\mathrm{b}$ is not fixed% as in \citep{DelPopolo2012b,DelPopolo2014}
. Instead, 
we assume that the system initially starts with %has 
a baryon fraction %that is
equal to the ``universal'' baryon fraction $f_\mathrm{b}=0.17 \pm 0.01$
\citep{Komatsu2009} while the final baryon fraction is calculated taking into account the major baryonic components: Intra-cluster Medium (ICM) (mainly primordial), intracluster stars, and stars in galaxies, (the latter%st
) determined from the star-formation processes (see %Appendix 
\ref{sec:model}). 
%
%The final baryon fraction, obtained in this way, has been compared with the baryon and gas fractions of the %clusters in \citep{Giodini2009} and is found to be in agreement with their results. {\bf c'e' %Gonzalez......Non c'e' accordo con fgas di Donahue}
%
In the same way, the %The 
``random'' angular momentum is no longer a parameter modified to improve the agreement between the model and data, but has been fixed following \citep{AvilaReese1998,Ascasibar2004}. Given the cluster %s 
mass, and the total mass profile, the model reproduces the DM, stars, and gas profiles. 
A qualitative summary of this model (see also \ref{sec:model} for details) goes as follows.
In our model, the protostructure contains baryons and DM. After growing
in a linear way, the density contrast becomes large enough to stop the 
protostructure expansion with the Hubble flow, making it recollapse. DM
collapses first and baryons follow. Clumps are formed because of
radiative processes, and collapse to the protostructure center to form
stars. At high redshifts ($ z \simeq 5$ in the case of a protostructure
of $10^9 M_{\odot}$), the collapsing DM compresses baryons (adiabatic
contraction). The formed clumps transfer energy and angular momentum to
DM through dynamical friction. Then the amplitude of DM random motions
increases, and DM moves toward the outskirts of the protostructure, 
resulting in a reduction of the central DM density of the forming
structure and erasing or flattening of the initial cuspy
profile. Protostructures giving rise to rotation supported galaxies
suffer from a further flattening due to the acquisition of angular momentum in
the collapse phase. SN explosions at a later epoch ($z \simeq 2$)
produce expulsion of gas, and the smallest clumps remaining after star
formation are disrupted \citep{Nipoti2015}. AGN feedback has a similar
effect on larger scales to that of SN feedback \citep{Martizzi2012}.  

%The results of our model were compared to the CLASH data as follows.  

\section{ Results and discussion}\label{sec:Results}

In \citep{DelPopolo2009,DelPopolo2012a,DelPopolo2012b,DelPopolo2014}, we
showed how the environment, angular momentum, and baryon content
influence the characteristics of cluster structure.  
The inner density profiles of clusters are flatter in clusters 
with larger angular momentum \citep{AvilaReese1998,Subramanian2000,Nusser2001,
Hiotelis2002,LeDelliou2003,Ascasibar2004,Williams2004,Ascasibar2006} 
%AvilaReese2001
and larger baryon fraction (especially in the central region). In
\citep{DelPopolo2014}, we reproduced the total and DM density profiles
of \citep{Newman2013a,Newman2013b} and those correlations found by these
authors.

The results obtained by  \citep{Newman2013a,Newman2013b} are based on
strong-lensing, weak-lensing, and improved stellar kinematics with
respect to their previous work \citep{Sand2004}.
The \citep{Newman2013b} data reduced the degeneracy
between the stellar and DM masses, thanks to the determination of the
stellar mass scale and by accounting for the BCGs homogeneity
\citep[][see their Sec.~4]{Newman2013b}.
This resulted in a more physically consistent analysis.
They showed that the total density profile is well approximated by a
cuspy NFW profile, while the DM profile is found to be flatter.  
 
% .......

\begin{figure} %[!tph]
  \begin{center}
    \begin{adjustwidth}{0.05cm}{0.05cm}
    \hspace{-2.85cm} \includegraphics[width=6.8in,height=5.8in]{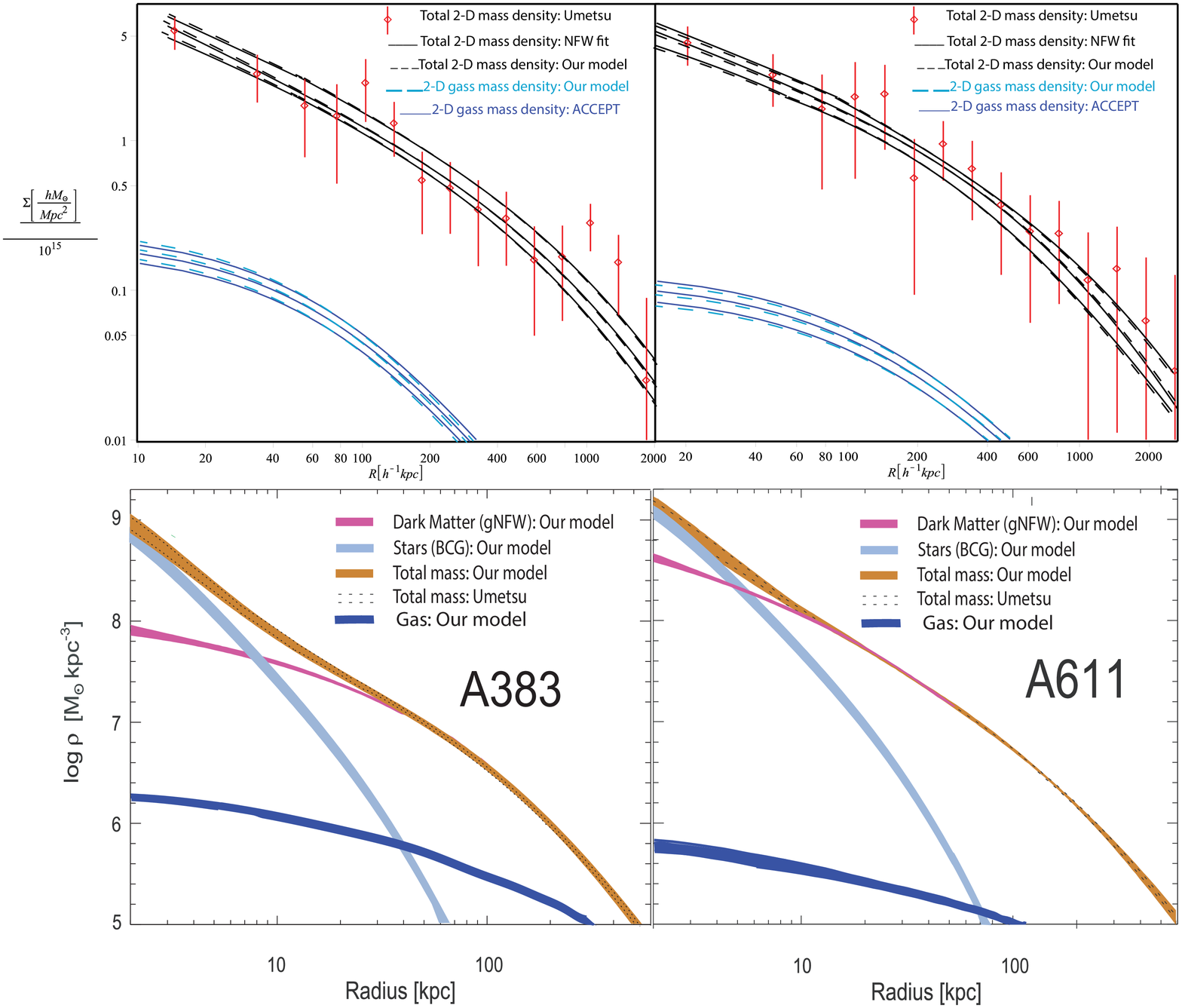}
%    \hspace{-2.85cm} \includegraphics[width=6.8in,height=5.8in]{fig1uuuu10.eps} 
%  \hspace{-2.85cm} \includegraphics[width=6.8in,height=5.8in]{fig1uuuu9.eps}   
%     \hspace{-2.85cm} \includegraphics[width=6.8in,height=5.8in]{fig1uuuu.eps}
    \end{adjustwidth}
\caption{ Total, and gas surface 
mass density profiles for A383 (top-left) and A611
 (top-right). The diamond symbols with error bars show the
 data from \citep{Umetsu2016}, while the central and external lines
 represent the best-fit NFW profile and the 68\% CL, respectively. The dashed lines are the result of our model.
  In the bottom parts of the same panels is plotted the gas surface density obtained projecting the 3-D surface density \citep{Baltz2009} {%\bf 
  given in the webpage of the ACCEPT project} \url{https://web.pa.msu.edu/astro/MC2/accept/}, solid lines, and the prediction of our model, dashed lines.
  The bottom-left and bottom-right panels show the 3D density
 profiles of A383 and A611 obtained with our model, respectively. The cyan, magenta,
 brown, and blue bands represent the stellar, DM, total matter, and gas density profiles,
 respectively. The dashed line represents the \citep{Umetsu2016} total density profile.
The width of the band indicates the $1\sigma$ uncertainty \citep[see
 Sec.~4.3 of][]{Newman2013b}.
%The arrows at the bottom represent the three-dimensional BCG half-light radius, $r_{\rm h}=1.34 R_{\rm e}$.
 }\label{fig:1}
%\label{resfig}
\end{center}
\end{figure}

In Fig.~\ref{fig:1} (top panels), we compare the total surface mass
density profiles $\Sigma$ obtained by \citep{Umetsu2016} (diamond
symbols with error bars, %and 
its NFW profile fit and the 68\% CL -- Confidence Level) with our model predictions, shown with dashed lines, 
%(the three lines, representing the best-fit NFW profile and the 68\% CL), 
for the cases of
A383 (left panel) and A611 (right panel).
As shown here, our model reproduces well the \citep{Umetsu2016} surface
mass density profiles and shows that, similarly to what was found in
\citep{Newman2013a,Newman2013b,DelPopolo2014}, the total mass profile is
well approximated by a cuspy NFW-like profile for both clusters. 
In the bottom parts of the same panels, the solid lines represent the gas surface density, (average value and the 68\% CL), obtained by projecting the 3-D 
%\Mov{\it [its is either 3D or surface but not both, right? same in caption]}%surface 
density \citep{Baltz2009} {%\bf 
obtained from the ACCEPT project} given in \url{https://web.pa.msu.edu/astro/MC2/accept/}. The dashed lines are the prediction of our model.

In the bottom panels, we show the 3D density profiles of A383 and A611 obtained
with our semi-analytical model. The cyan band represents the stellar
content, the magenta one is the DM content, the brown one is the total mass,  and the blue band the gas.
The dashed lines represent the total mass from the \citep{Umetsu2016} data. 
%{\bf 
%and the stellar mass from \cite{Newman2013b}. 
%{\bf
%  The DM was not shown since it is given by the difference between the total, and stellar mass.

%
%The plots show that our model reproduces well all mass components.
%
%}
%TOLTO
%Here we stress that \citet{Umetsu2016} lensing analysis just gives the total mass, and no information is %obtained on the stellar component. Accordingly, in the top panel we only plotted the result of the 
%\citet{Umetsu2016} lensing analysis and, since baryonic distribution was not obtained by \citet{Umetsu2016}, %we have no way to compare it with the model results. 
%In order to show that the trends and correlations obtained in the next figures are "required" by data, we %somehow need to show that the model is not only able to fit the total mass but also the stellar (baryonic) %mass. To this aim, in the bottom panels, we compared the result of our model with the stellar distribution %of two clusters out of the three clusters that are common between CLASH clusters and \citet{Newman2013b} %clusters. In fact, differently from the \citet{Umetsu2016} analysis, \citet{Newman2013b} gives the stellar 
%(baryonic) distribution, making possible a comparison between the stellar (baryonic) content predicted by %the model and that observed.}
%

\begin{figure} %[!tph]
\begin{center}  
    \begin{adjustwidth}{0.05cm}{0.05cm}
     \hspace{-2.85cm}\includegraphics[width=6.5in,height=4.2in]{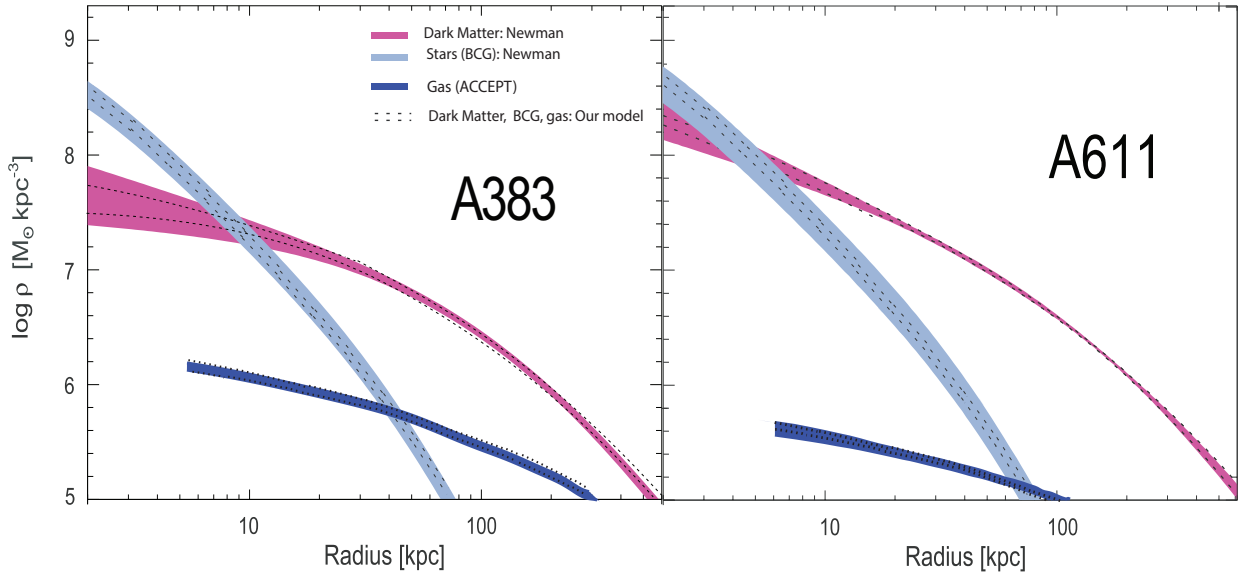}
%     \hspace{-2.85cm}\includegraphics[width=6.5in,height=4.2in]{fig2Newm3_2.eps}
%     \hspace{-2.85cm}\includegraphics[width=6.5in,height=4.2in]{fig2Newm3_1.eps}
%      \hspace{-2.85cm}\includegraphics[width=6.5in,height=3in]{fig2Newm3.eps}
             \end{adjustwidth}
\caption{ Comparison of the \citep{Newman2013b} DM, stars, and gas density profiles of A383 and A611 with our model's predictions. The cyan, magenta, and blue bands represent the stellar, DM, and gas density profiles, respectively. The width of the band indicates the $1\sigma$ uncertainty \citep[see
 Sec.~4.3 of][]{Newman2013b}. 
 %The arrows at the bottom represent the three-dimensional BCG half-light radius, $r_{\rm h}=1.34 R_{\rm e}
 %$.}
 }
 \label{fig:newm}
%\label{resfig}
\end{center}
\end{figure}

The arrows at the bottom represent the three-dimensional BCG half-light radius.
%, $r_{\rm h}=1.34 R_{\rm e}$.
%The line segment indicates the slope of the NFW profile. 
In each case, the width of the band indicates the $1\sigma$ uncertainty \citep[see Sec.~4.3 of][]{Newman2013b}.

\begin{figure} %[!tph]
\begin{center}
\includegraphics[width=4.8in,height=3.8in]{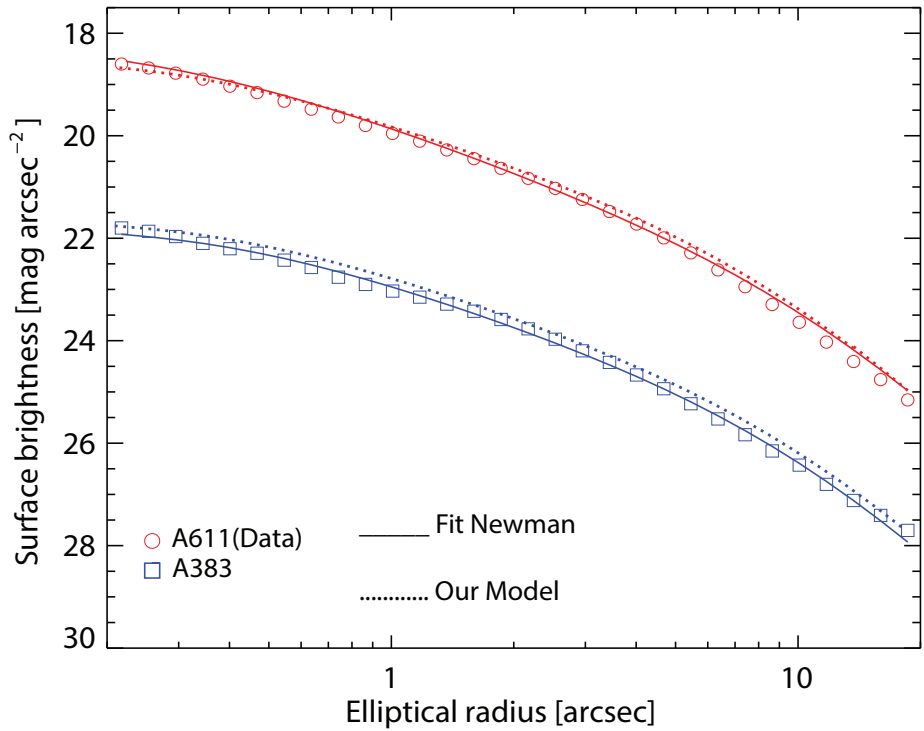}
\caption{
Comparison of the \citep{Newman2013b} profiles of surface brightness for the BCGs of A611, and A383
with our model's predictions. The symbols and the solid lines  are the data, and pseudo-isothermal elliptical mass distribution \citep[dPIE, see Eq. 1 of][]{Newman2013b} fits to the data, respectively, obtained by \cite{Newman2013b}. The dotted lines are the dPIE fits done in our model.  
%The critical interval is estimated approximately as where the surface
%brightness exceeds 10% of that at the outer limit of the kinematic data (indicated by top arrows).
}\label{fig:1a}
%\label{resfig}
\end{center}
\end{figure}

%{\bf (KU: I moved up the following paragraph, which describes our
%determinations of the 2D and 3D density slopes, because
%these quantities are
%discussed in subsequent paragraphs along with figures)}
Tables \ref{tab:1} and \ref{tab:2} summarize all the CLASH parameters
found in our analysis.

\begin{figure} %[!tph]
\begin{center}  
    \begin{adjustwidth}{0.05cm}{0.05cm}
\includegraphics[width=5.8in,height=5.8in]{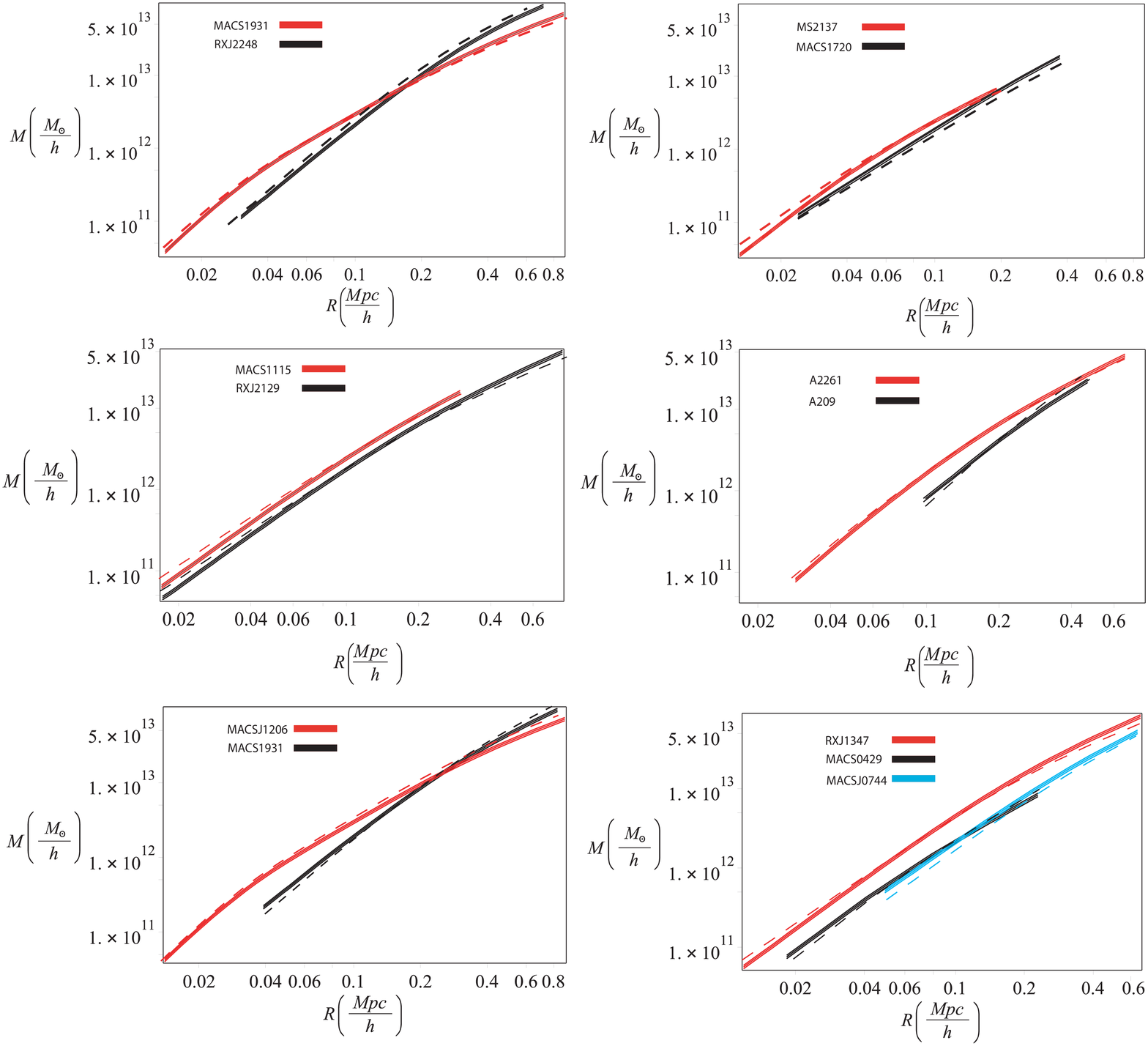}
             \end{adjustwidth}

\caption{Comparison of the gas mass profiles inferred by JACO \Mov{\citep[Joint Analysis of Cluster Observations,][]{Mahdavi2007}} from Chandra, solid thick lines, with the predictions of our model, dashed lines, for all the clusters except A383, and A611, since for those clusters the comparison was already done in Figs.~\ref{fig:1},\ref{fig:newm}% 1, 2
.}\label{fig:gas}
%\label{resfig}
\end{center}
\end{figure}

The 2D density profile slope, $\alpha_\mathrm{2D}$, was obtained using 3 
adjacent radial bins fitted with a 2-parameter power-law profile,
accounting for the averaging effect \citep{Umetsu2016}.\footnote{The surface mass density profile is described by 15 radial bins. Thus the local slopes can be measured at 13 sets of 3
adjacent radial bins, yielding the logarithmic density slope as a
function of the cluster-centric radius.}
The 3D DM density slope $\alpha$ is obtained from fitting the spherical
DM profile 
%({\bf KU: spherical profile $\to$ spherical DM profile, correct?})
with a generalized NFW profile (gNFW).\footnote{The gNFW profile is written as
\begin{equation}
\rho_\mathrm{DM}(r)=\frac{\rho_\mathrm{s}}{(r/r_\mathrm{s})^\alpha (1+r/r_\mathrm{s})^{3-\alpha}},
\end{equation}
with a central cusp slope given by $d \log{\rho_\mathrm{DM}}/d \log{ r} \rightarrow -\alpha $ for $r \rightarrow 0$}

\begin{figure} %[!tph]
\begin{center}
\includegraphics[width=5.8in,height=3.8in]{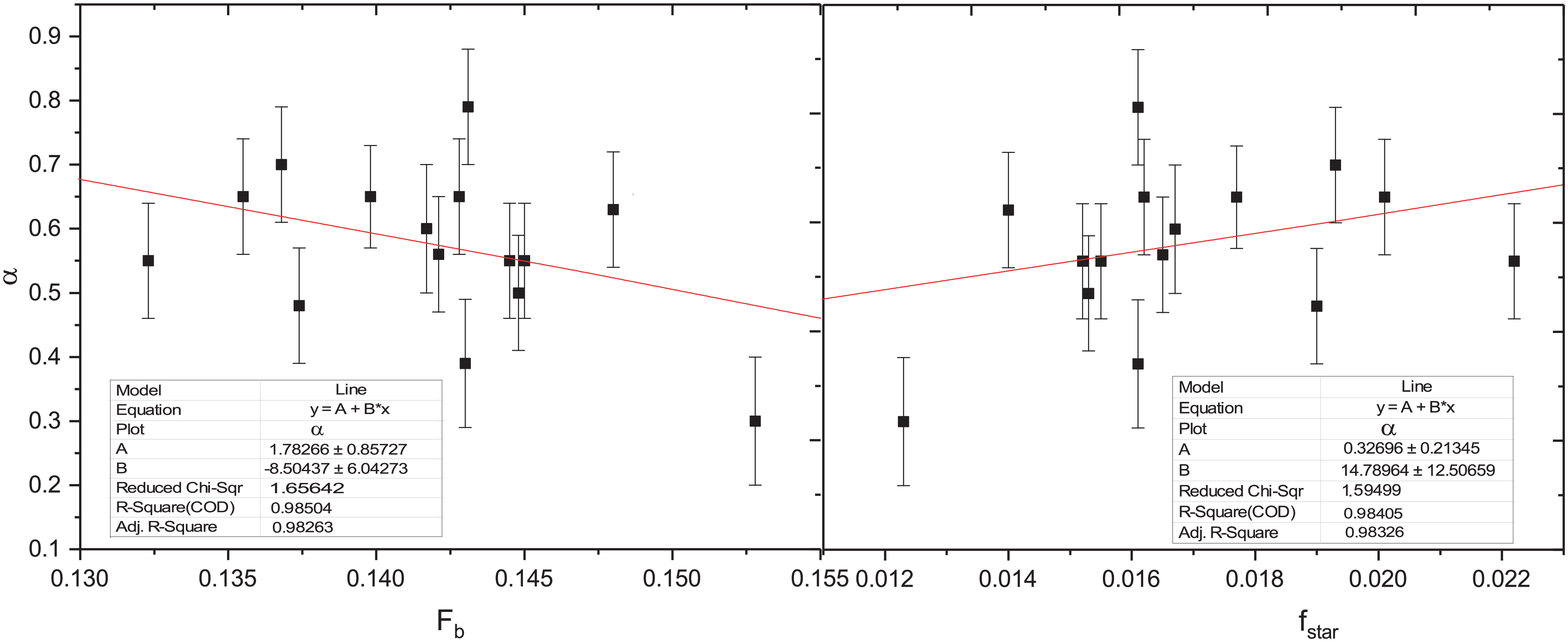}
\caption{Inner 3D slope of the DM density $\alpha$ versus cluster baryon fraction
 $F_\mathrm{b}$ (left), and 
 $\alpha$ versus stellar mass fraction $f_\mathrm{star}$ (right),
 obtained with our model (square symbols with error bars).
In each panel, the solid line represents the orthogonal distance regression (ODR) fit to the data. 
%SPEARMAN: -0.26274; 0.25   
The Spearman correlation coefficients for the respective relations are 0.43 and -0.44.}\label{fig:2}
%\label{resfig}
\end{center}
\end{figure}

In the inner $\simeq 5-10$\,kpc region \citep[see][Fig.~3]{Newman2013b}
of A383 and A611, where the BCG mass becomes comparable or larger than
the DM mass, Fig.~\ref{fig:1} (bottom panels) shows that the density
profiles flatten (similarly for the other clusters, whose plots were
not shown). 

%{\bf
  At this point, we have to notice that while the flattening of the DM distribution occurs within $ \simeq 10$ kpc, region in which the mass of the BCG is dominating, the top row of Fig. 1 shows that the surface density does not extend to radii $<10$ kpc.  

%
%\textcolor{red}{Scrivere che il modello e' testato ecc, aggiungere i successi, citare i grafici che vuole il %referee, dire che in ogni caso il nostro articolo da' delle previsioni che saranno verificate in futuro }    %\Mov{\bf [is this what is addressed below?]}
% 
  
%How can they constrain the DM profile at such radii
In that region, we reproduced the total density profile given by \citep{Umetsu2016} with our model, as the model produces the three dimensional distribution of total mass, DM, and baryonic density profiles. 
%In other words, 
In the same way as the model reproduces the total mass density profile of \citep{Umetsu2016}, and “reconstructs” the DM, and baryonic distribution, it also reconstructs the density profiles a radii $< 10$ kpc. 
%}
 In other terms the model gives a "physical extrapolation%estrapolation
" in the region for which there is no data. A legitimate question one can ask is whether the model is robust enough %how robust is 
%the model 
to allow this kind of extrapolation%estrapolation
. 
The model, similarly to hydro-dynamic simulations \citep[see][]{Martizzi2012,Martizzi2012a, Martizzi2013}, takes into account a large number of physical effects (e.g., gas cooling, star formation, supernovae feedback, AGN feedback, etc.). It %The model 
has been checked in several different situations. In %and, in 
some studies, it has predicted several important effects in advance of hydro-dynamical simulations, and %it 
has also been compared with hydro-simulations. An extensive discussion on the robustness of the model can be found in the ending part of the %Appendix 
\ref{sec:model}.

\begin{figure} %[!tph]
\begin{center}
\includegraphics[width=5.8in,height=3.8in]{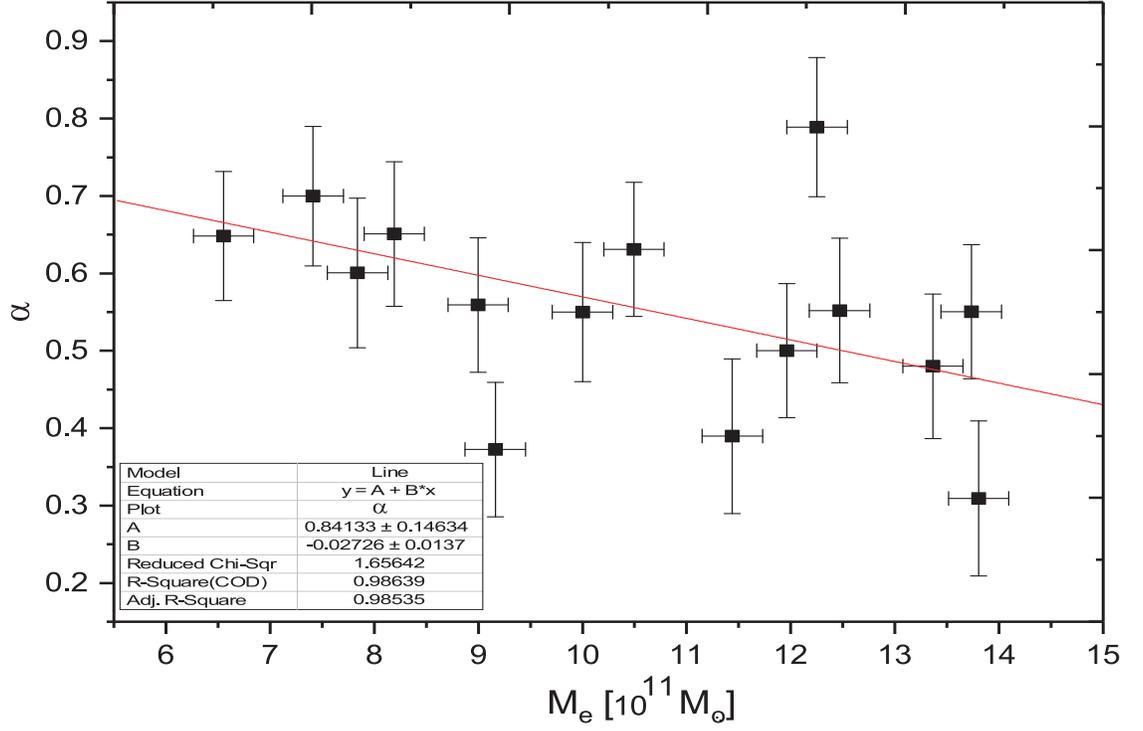}
\caption{Correlation between the inner DM slope $\alpha$ and the BCG
 mass $M_\mathrm{e}$. The BCG stellar mass $M_\mathrm{e}$ was obtained using our
 model. For each cluster, the uncertainty of the BCG mass is assumed to
 be 0.07\,dex \citep{Newman2013b,DelPopolo2014}.
The solid line is the ODR fit to the data. The Spearman correlation coefficient is -0.6.
%{\bf KU: Please correct the figure for the xlabel}
}\label{fig:3}
%\label{resfig}
\end{center}
\end{figure}

\begin{figure} %[!tph]
\begin{center}
\includegraphics[width=5.8in,height=3.8in]{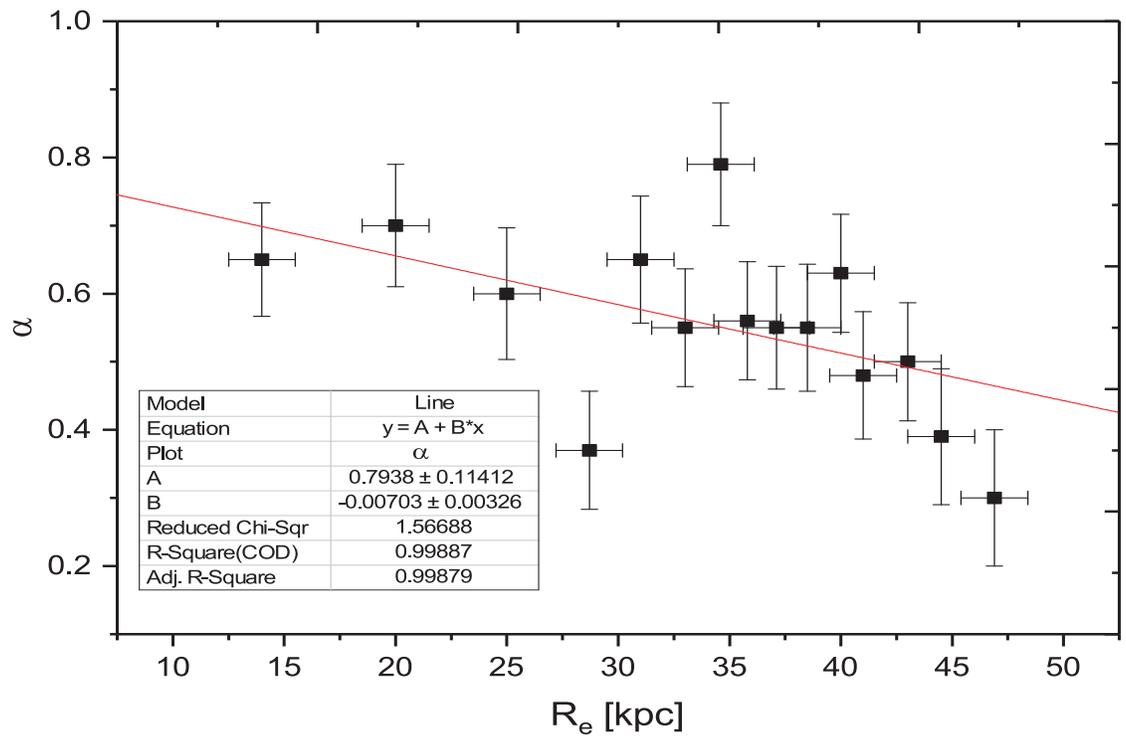}
\caption{Correlation between the inner DM slope $\alpha$ and the BCG
 effective radius $R_\mathrm{e}$.
The line represents the ODR fit to the data, with a Spearman correlation
 coefficient of $-0.63$ and a $p$-value of $0.0012$.}\label{fig:4}
%\label{resfig}
\end{center}
\end{figure}

In addition, %We want also to add that 
the correlations we find in the paper are related to properties of the BCG (Figs. 6-8). In Figs.~\ref{fig:3}, \ref{fig:4}% 6, 7
, we show a correlation between the inner slope of DM and 
parameters of the BCG %(Figs. 6-7), 
which, as discussed by \citep{Newman2013a,Newman2013b}, is very little influenced by the gas ($\frac{\Delta dlog \rho}{d log r} \leq 0.05$). Similarly in Fig.~\ref{fig:6}, %9 
the correlation is found %is 
between the mass content within %in 
100 kpc, mainly composed of DM, and the mass inside 5 kpc, dominated by the BCG mass. 

However, as previously discussed, in the CLASH sample there is no data inside 10 kpc. Moreover, as we discuss in the following, there exists only %is just 
one study on CLASH BCGs \citep{Burke2015} in the literature. Nevertheless, that study %, which however 
gives discrepant results compared with \citep{Newman2013a,Newman2013b} for the characteristics of the BCGs of the clusters common to the two studies \citep[i.e.][]{Newman2013a,Newman2013b,Burke2015}, namely A383, and A611, while \citep{Newman2013a,Newman2013b} are in agreement with our BCGs characteristics. 
{%\bf 
Therefore, we prefer not to compare the correlations we found from the model predictions with \citep{Burke2015} data
% we prefer not to compare our data with that of \citep{Burke2015} 
because of the apparent tension with other studies.}
%
%Therefore,  the correlations we found from the model predictions, cannot be compared with observations, %since there are no CLASH data for the BCGs (nor inside 10 kpc). 
%
Our results are then based on the model robustness (see the following) and should be checked against future observations.

The increased role of the baryon mass at these radii (mainly the BCG mass) steepens the total density profile compared with the DM density profile, whereas for radii $\geq 5-10$\,kpc, in all
clusters, DM dominates over the baryon component. As a result,
the outer total and DM density profiles
are very similar, and their slopes outside the inner 
region are comparable in the different clusters, in agreement with the
NFW profile. Since the total density profile is consistent with the NFW
profile at $r \simeq 5- 30$\,kpc,
this implies a ``tight coordination'' between the stellar distribution
and the inner DM profile, as found by \citep{Newman2013b}.
Since the total mass (DM and baryonic matter) in the inner cluster
region follows the NFW form, while the baryonic 
component is dominant in the $5-10$\,kpc central region, the DM 
central density profile is flatter than the NFW profile.

These trends are more clearly visible in Fig.~\ref{fig:2} and better in
Figs.~\ref{fig:3} and \ref{fig:4}.

Although the model, summarized in points a-e of the \ref{sec:model}, was validated in previous papers, we compare, in Fig. \ref{fig:newm}, the model's prediction for the DM, and baryons (stars) density profile with the observations of \citep{Newman2013b}, as an illustration of how the model equally correctly predicts the DM and baryons distribution. For consistency with Fig.~\ref{fig:1}, we only display the results for clusters A383 and A611. However, all \citep{Newman2013b} clusters were checked. 

 As seen in Fig. \ref{fig:newm}, our model (dashed lines) gives a very good approximation to the DM density profile (magenta band), baryons (i.e. stars, %stars) (
cyan band), and gas density profile. 
We provide a further check on our model's baryon distribution predictions in Fig.~\ref{fig:1a}, with a comparison, for the two illustrative clusters, of the model with the observed \cite{Newman2013b} surface brightness profiles.
They both present good agreement between observed and model surface brightness.%} 
 
For those clusters%}
, %finding a good agreement.
%{\bf
their %}
%The
stellar mass-to-light ratio, $M_*/L$, is equal to: 
%2.03 (MS2137), 
2.3 (A383), and 2.2 (A611), in agreement with \cite[Table 4]{Newman2013b}.
%}

With %In 
the same aim to validate %optic of validating 
the model, in Fig.~\ref{fig:gas} %, 
we compare the gas mass profile, 
$M_{\rm gas}$. This %, which 
is one of the best proxy to check the SAM predictions, because it can be directly recovered from X-ray observations, without %almost 
any assumption. We present profiles %did it 
for all clusters (except A611, and A383 for which we already checked the density profile in Fig.~\ref{fig:1}% 1
) obtained from {%\bf 
\citep[$Chandra$/JACO dataset]{Donahue2014}\footnote{ \rm %\bf 
JACO \citep{Mahdavi2007} is a tool to derive gas and hydrostatic equilibrium density profiles.}}
 with the profiles %those obtained 
from our model. As shown, the plots display %shows 
a good agreement between the observations (average and 68\% CL) and %with 
the average values of the SAM's prediction. A comparison between the stellar content of the clusters and our model is not possible because data are lacking. Using the profiles of \citep{Lin2004}, or calculating $M_{\rm star}$, assuming that all clusters have the same baryon fraction, $f_{\rm bar}$, as in % as 
$M_{\rm star}=f_{\rm bar} M_{Total}-M_{\rm gas}$, gives average profiles of $M_{\rm star}$. To correctly compare with our predictions%have a correct comparison
, we need measurements of the stellar component which are not yet available%we do not have
. A similar problem is present for the Brightest Cluster Galaxy (BCG).
% questo e’ un modo per validare il tuo modello nella sua produzione relativa di barioni; nota pero’ che ti %dara’ un Mstar “medio” visto che imponi la baryon fraction “fb” che difficilemnte puoi usare per fare le tue %stime di correzioni che invece necessitano non di valori “medi” ma di misure “dirette” (e.g. da cinematica %stellare nella BCG)
%M_DM= M_WL-Mgas- Mstar
We can %do comparisons 
only compare %for 
clusters in which the stellar component and the BCG characteristics have been measured. Such comparison was done for clusters with the corresponding measurements%We have done this for clusters for which the quoted quantities have been measured
: A611, and A383 plotted in Fig.~\ref{fig:newm}% 2
\footnote{Although not presented%Even if not plotted
, we %also 
get a similar result for MS2137}. 
Recall %We want to recall 
that in previous papers we showed
%, using our model, 
how our %the 
model gives good predictions for the gas, stars, and DM content in structures \citep[e.g.][]{DelPopolo2018}, as well as the baryonic Tully-Fisher (TF)
%{\it [cite?]}}%TF 
relation (i.e., baryonic mass, rotation velocity relation), the $M_{\rm star}-V_{\rm rot}$, $M_{\rm star}-\sigma$ relations (for %, being 
$\sigma$ the dispersion velocity and $V_{\rm rot}$ the circular rotation velocity) %, 
and the $M_{\rm star}-M_{\rm halo}$ relation \citep[see][]{DelPopoloPace2016}.

In Fig.~\ref{fig:2} we show the correlation between the 3D DM density
slope $\alpha$ and the baryon fraction $F_\mathrm{b}$  
(left panel) and the correlation between $\alpha$ and the stellar mass
fraction $f_\mathrm{star}$ (right panel).
The Spearman correlation coefficients are 
$\simeq 0.43$ and $\simeq -0.44$, respectively.
We note that these correlations are stronger than those of the 2D slope
of the total density ($\alpha_\mathrm{2D}$) with $F_\mathrm{b}$
and $f_\mathrm{star}$ (not shown). This is because, unlike the 2D
total mass  profile, the DM density profile varies significantly from
cluster to cluster 
%({\bf KU: If the DM density profile varies from cluster to cluster, implying a large scatter, isn't it that %the correlation is ``weaker'' in DM?}). 

Figure \ref{fig:3} compares the inner 3D DM slope ($\alpha$) and the BCG
stellar mass ($M_\mathrm{e}$).  
The DM slope parameter $\alpha$ was obtained by parameterizing each
cluster with a gNFW profile. 
From the figure, it is evident that the larger the cluster's BCG mass,
the flatter its inner density profile, in agreement with
\citep{DelPopolo2012b,Newman2013b,DelPopolo2014}. This is expected for
the following reason: Since the total (DM + baryons) density profile is well
represented by a cuspy NFW profile and is
dominated by stellar baryons inside $5-10$\,kpc, 
this implies that the steeper the baryon inner profile, the flatter the
DM profile. The best-fit line is derived from orthogonal distance
regression (ODR), which takes into account the uncertainties in the
variables different from the least-square method.
%
%As in \citep{Newman2013b,DelPopolo2014}, the error on the BCG mass was
%assumed equal to 0.07 dex. 
%
We calculated again the Spearman correlation coefficient, which 
in this case is $-0.6$, with a $p$-value of $0.02$,\footnote{The
probability that the ``null'' hypothesis (the true correlation is zero) is true.}
testifying for the correlation between $\alpha$ and $M_\mathrm{e}$.

Here, we cannot make a statistical comparison between the parameters
derived from our modeling and those measured directly from observations,
because there is just one systematic study of CLASH BCGs
\citep{Burke2015} in the literature,  
%outside the CLASH collaboration, 
and it does not investigate the $\alpha$--$M_\mathrm{e}$
correlation. The $\alpha$--$M_\mathrm{e}$ relation was studied 
%{\bf
for two of the CLASH clusters, namely A383, and A611 by %}
\citep{Newman2013b,DelPopolo2014}.
Our results for the $\alpha$--$M_\mathrm{e}$ relation
are in agreement with \citep{Newman2013b}.
Here, we would like to point out that
a comparison of the BCG masses common to \citep{Burke2015} and
\citep{Newman2013b,DelPopolo2014} %{\bf
  (A383, A611) %}
 shows that while
those in \citep{DelPopolo2014} are in agreement with
\citep{Newman2013b}, %{\bf
only A383 agree with \citep{Burke2015}%}
.

In Fig.~\ref{fig:4}, we show the $\alpha$--$R_\mathrm{e}$ relation,
with $R_\mathrm{e}$ the BCG effective radius (radius containing half of
the total light).
The figure shows that a cluster with larger $R_\mathrm{e}$ have a
shallower inner slope, which is in line with what was discussed above.
That is, clusters with larger BCGs contain more stellar baryons and thus
less DM 
($M_\mathrm{DM}=M_\mathrm{total}-M_\mathrm{b}$)
in their central region, resulting in a flatter DM slope.
The solid line in the figure represents the ODR fit,
with  Spearman correlation coefficient of
$-0.63$ and a $p$-value of $0.0012$. 
%({\bf KU: 0.001159 $\to$ wedon't need so many significant digits}).
% 0.001159

%
%{\bf We want to stress that both in N13b, and the present paper, the correlation between the core radius, 
%$r_{core}$, and the BCG effective radius, $R_e$, is strongly dependent on the two data points at largest %radii. }
%

\begin{figure} %[!tph]
\begin{center}
\includegraphics[width=5.8in,height=3.8in]{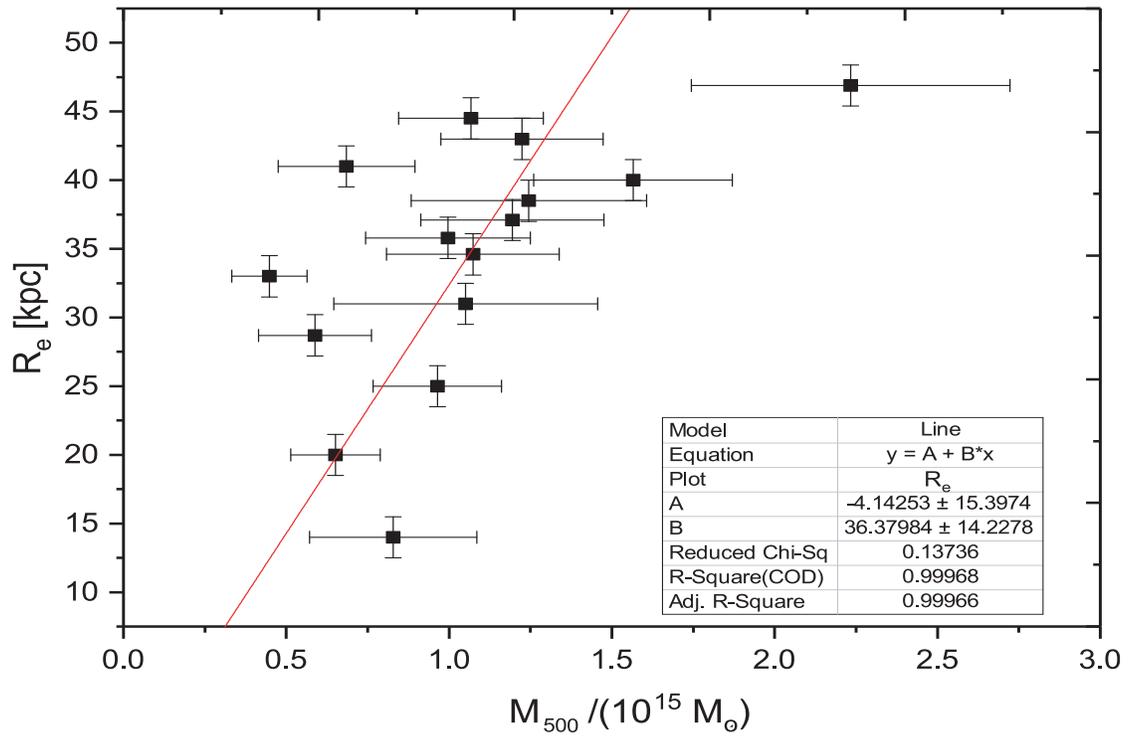}
\caption{Correlation between the BCG effective radius $R_e$ and the
 cluster halo mass $M_{500}$. The solid line shows the ODR fit, with 
a Spearman correlation coefficient of $-0.68$ and a $p$-value
 of $0.005$. }\label{fig:5}
%\label{resfig}
\end{center}
\end{figure}

\begin{figure} %[!tph]
\begin{center}
\includegraphics[width=5.8in,height=3.8in]{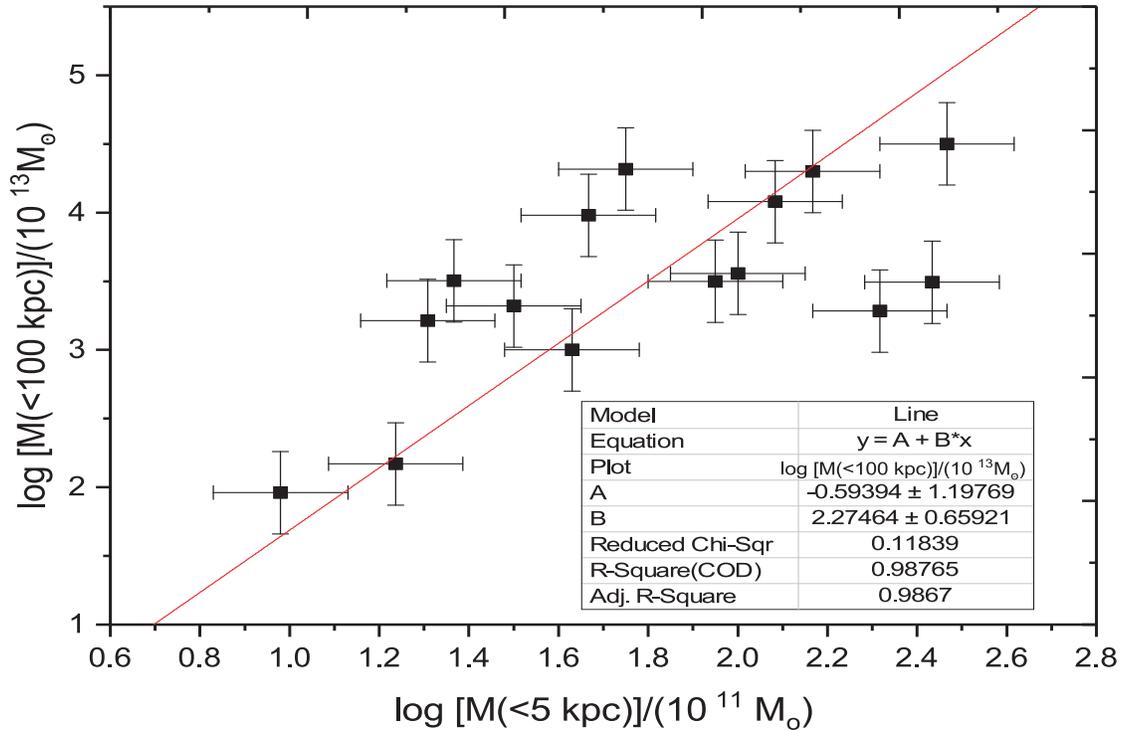}
\caption{Total cluster mass contained within 100\,kpc (dominated by DM) versus that
 contained within 5\,kpc (dominated by stellar baryons). The solid line
 is the ODR fit with a Spearman correlation coefficient of $0.64$ and a
 $p$-value of $0.011$.}\label{fig:6}
%\label{resfig}
\end{center}
\end{figure}

In Fig.~\ref{fig:5}, we show the correlation between $R_\mathrm{e}$ and
the halo mass $M_{500}$ (at $r_{500}\sim 1.4$\,Mpc for our sample). 
The solid line represents again the ODR fit. This comparison exhibits a
positive correlation with a Spearman correlation coefficient of
$-0.68$ and a $p$-value of $0.005$. This
is in agreement with previous studies  \citep[e.g.,][]{Kravtsov2013}, but
in contradiction again with \citep{Burke2015}. The latter found no
correlation between the BCG mass and the cluster halo mass.
They ascribed this lack of correlation to a selection bias in the CLASH
sample (whereby clusters with BCGs in a narrow mass range have been
selected), but without a clear justification.
%due to
%small mass BCGs selection of clusters.
However, as previously noticed, the BCG mass estimates between
\citep{Burke2015} and \citep{Newman2013b} in their overlapping clusters
are in striking conflict with each other. Moreover, the inclusion of the
high-magnification CLASH clusters in \citep{Burke2015} may imply a
bias in their results, because they are highly disturbed merging clusters.

Our model \citep{DelPopolo2009,DelPopolo2012a} can explain the resulting
density profiles (Fig.~\ref{fig:1}) and correlations (Figs.~\ref{fig:3}
and \ref{fig:4}) as follows:
The DM protostructure starts to collapse at high $z$ (linear phase),
forming potential wells for baryons to fall in. In their collapse,
baryons radiate away part of their energy and form clumps, condensing into
stars \citep[see][Secs.~2.2.2 and 2.2.3]{Li2010}, while compressing DM
\citep[``adiabatic contraction'',][]{Blumenthal1986,Gnedin2004}.
At around $z\geq 5$ \citep[][see Figs.~3 and 5 therein]{DelPopolo2009},
this process dissipationally produces a steep density profile, the main
structure of the BCG \citep[see also][]{Immeli2004,Lackner2010} with
scale radius
$R_\mathrm{e} \simeq 30$\,kpc, similar to size-scales of high-redshift
massive galaxies \citep{Trujillo2006,Newman2013a,Newman2013b}. Extra
stars are added in the outer regions by satellite mergers onto the
proto-BCG \citep[e.g.,][]{Naab2009,Laporte2012}

Moreover, dynamical friction from DM particles induces orbital
decay of the baryon clumps.
As a result, DM particles move outward as the baryon clumps move
toward the cluster center, thus reducing the central DM density
\citep{ElZant2001,ElZant2004,Nipoti2004,DelPopolo2009,Cole2011,Inoue2011,Nipoti2015}. 

AGN feedback has been proposed as an alternate mechanism to flatten the
DM profile \citep[e.g.,][]{Martizzi2012}. However, it appears to be
``too effective'' because the 10\,kpc core produced is much larger than
what is observed \citep{Postman2012}.  

Our physical model thus predicts, at the same time,
a flattening of the inner DM density profile and an anti-correlation
between the inner DM slope $\alpha$ and the cluster's central baryon content
\citep{Nipoti2004,DelPopolo2009,DelPopolo2012a}.
This is because the density profile shape is primarily the result of
interaction between DM and baryons clumps through dynamical
friction and subdominantly of the action of SN/AGN feedback.

%
%a) angular momentum, b) baryonic fraction, c) virial mass, d) Supernovae and AGN feedback.
%

Dissipative baryonic formation of the proto-BCG ($z \geq 2$) follows in
our model, where stars merge onto the BCG
\citep[e.g.,][]{Naab2009,Laporte2012} and satellites infall toward the
cluster center,
thus kinematically ``heating'' DM and flattening the inner DM slope
\citep{ElZant2001,ElZant2004,DelPopolo2009,Cole2011}.

A further correlation is expected between the star-dominated, innermost 
  cluster mass and the cluster mass within a region that is already
  established before the BCG formation and thus almost unchanged
  subsequently \citep{Gao2004}.
  This has been examined in Fig.~\ref{fig:6} by comparing the
  the cluster mass within 5\,kpc
  (composed mainly of stars) and that within 100\,kpc (mainly of DM) for
  our CLASH sample.
  We find a Spearman correlation coefficient of $0.64$ and a
  $p$-value of $0.011$.  
Such a strong correlation implies (see also Sec.~\ref{sec:Conclusions})
that the cluster's progenitor halo and the innermost BCG region were
formed at higher redshifts, and they have been subject to little
  evolution subsequently, whereas the cluster outskirts have grown via
  the secondary mass accretion. This ``inside-out'' growth scenario of
  cluster formation has been proposed in the framework of $\Lambda$CDM 
\citep[e.g.,][]{Wechsler2002,Zhao2003} and recently confirmed from a
joint lensing and X-ray analysis of the cluster-halo fundamental plane
for the CLASH sample
\citep{Fujita2018a,Fujita2018b}. 
%({\bf KU: I've expanded the discussion in the context of the inside-out growth scenario and our recent %discovery of the cluster-halo fundamental plane})

These correlations confirm the central role of baryons in shaping the DM
density profile in galaxy clusters, in particular for the central $\lesssim
10$\,kpc region. The BCG characteristics correlate with baryonic and
cluster masses. The total density profile agrees well with the NFW form,
while the DM density profile exhibits a shallower inner slope with
$\alpha<1$.

Our findings
are consistent with \citep{Sand2004,Newman2011} on the %{\bf
cluster A383%}
. Our results are also in line with
\citep[][]{Zappacosta2006}, who found that the
innermost region of the (non-CLASH) cluster A2589
has a shallow DM profile assuming reasonable values of
mass-to-light ratios.
%%%%
%({\bf KU: You were comparing A611 and A1703
%\citep{Limousin2008,Richard2009}, two different clusters. Considering
%cluster-cluster scatter, this comparison is not useful}) 
%%%%
%Our DM density slope for A611 is similar to
%\citep{Limousin2008,Richard2009}'s gNFW halo+BCG fit, giving
%$\alpha_{\textrm{DM}} = 0.92^{+0.05}_{-0.04}$.
Our cored DM profiles have a mean slope of $\alpha=0.54 \pm 0.05$,
in agreement with
\citep{Martizzi2012,Martizzi2013} and \citep{Newman2013b}. 

DM-only simulations produce halos with cuspy, steeper density profiles.
In the Phoenix project, \citep{Gao2012} analyzed zoomed-in re-simulations of
cluster-size halos drawn from a cosmologically representative volume in
$\Lambda$CDM, finding that the central density cusp (at their innermost
resolved radius of
$r\sim 2\times 10^{-3}r_{200}$)
has an average logarithmic slope of $\alpha\simeq 1.05$ with a halo-to-halo
scatter of
$\sim 20\%$. 
%({\bf KU:  I did not understand what you mean about 120 pc etc. Anyway you were referring to galaxy-scale %halos, which are irrelevant here.. I've rewritten it by referring to Gao+08}).
This level of ``diversity'' is apparently at odds with the
observed shallow slopes of the inner DM density.
%including the
%minimum slope at 120 pc, $\alpha \simeq -0.8$, and scatter of
%\citep{Stadel2009}.
%Although limits in the techniques (e.g., different
%dynamic ranges, BCG role, simplified modeling of clusters) can partially
%explain the scatter,  they don't alleviate the slope
%differences.
%%
Including the dominant central baryonic physics on top
of such DM simulations can reconcile the discrepancy and reproduce the
cored DM distribution, by accounting for dynamical friction from baryonic clumps
\citep{ElZant2001,ElZant2004,RomanoDiaz2008,DelPopolo2009,Cole2011} 
and
density flattening driven by SN/AGN feedback
\citep{Pontzen2014,Martizzi2012}. These mechanisms are described in Appendix \ref{sec:model}. 

On the other hand, more radical solutions have also been proposed:
alternative DM models \citep[e.g.,][]{Colin2000,SommerLarsen2001,Peebles2000,Kaplinghat2000},
modifications of the matter power spectrum at small scales
\citep[e.g.][]{Zentner2003}, and even modified gravity.

We find that the total density profile $\rho_\mathrm{tot}(r)$ of the
X-ray-selected CLASH sample is NFW-like, as found in previous studies
\citep{Umetsu2014,Umetsu2016,Umetsu2018}. 
The average slope of the total density for our sample is
$\langle \gamma_\mathrm{tot}\rangle=1.05 \pm 0.02$,
where 
$\gamma_\mathrm{tot}=-d\log{\rho_\mathrm{tot}}/{d \log{r}}$,
and the average slope is obtained by linear
fitting in the plane of $\log{r}$--$\log{\rho_\mathrm{tot}}$ in the
radial range
$r= (0.003-0.03) \times r_{200}$.\footnote{The BCG and the the DM halo
are distinct components of the cluster model, while
$\gamma_\mathrm{tot}$ is a derived, composite parameter.}
This is in agreement with collisionless DM simulations
($\gamma_\mathrm{tot}\simeq 1$), and in line with the finding of
\citep[][Sec.~9]{Newman2013a},
$\langle \gamma_{\textrm{tot}} \rangle = 1.16 \pm 0.05{}^{+0.05}_{-0.07}$
defined in the same range $r= (0.003-0.03)\times r_{200}$.

Observational results of the asymptotic total density profiles
also agree with our results (i.e., NFW-like at $r \gtrsim 5-10$\,kpc):
$\alpha_\mathrm{tot} = 0.96^{+0.31}_{-0.49}$ for MACSJ1206 \citep{Umetsu2012}
and
$\alpha_\mathrm{tot} = 1.08\pm 0.07$ for A383 \citep{Zitrin2011}.\footnote{It is important to stress that,
similarly to \citep{Newman2013a},
$\langle\gamma_\mathrm{tot}\rangle$ is the average slope of the total density
measured in the range $r= (0.003-0.03)\times r_{200}$
and is different from $\alpha_\mathrm{tot}$, which is the asymptotic
inner slope of the total density assuming the gNFW profile.}
Excluding the innermost 40\,kpc$h^{-1}$ region,
the stacked strong+weak lensing analysis of four ``superlens'' clusters 
(A1689, A1703, A370, Cl0024+17) gives
$\alpha_\mathrm{tot} =0.89^{+0.27}_{-0.39}$ \citep{Umetsu2011}.

\section{Conclusions}\label{sec:Conclusions}

In this paper, 
we have studied and characterized the total density profiles for a 
sample of 15 X-ray-selected CLASH clusters by improving our earlier
analysis \citep{DelPopolo2014} based on several clusters from
\citep{Newman2013a,Newman2013b}. The primary goal of this study was to
test the physical picture of cluster formation proposed by
\citep{Newman2013a,Newman2013b} in the frame work of a modified version
of the physical model developed by
\citep{DelPopolo2012b,DelPopolo2014}. 
To this end, we analyzed binned surface mass density profiles of \citep{Umetsu2016}
derived from their strong-lensing, weak-lensing shear
and magnification analysis of high-quality {\em HST} and Subaru data.
For each cluster, we extracted the radial profile of the total 3D density
assuming spherical symmetry. We have used our semi-analytical model to
interpret the total 3D density profile, which allows us to compute the
baryon density profile and thus the DM density profile for the cluster.
 
The total 3D mass density profile for our sample is characterized by a
logarithmic slope of
$\langle\gamma_\mathrm{tot}\rangle =1.05 \pm 0.02$ in the radial range
$r=(0.003-0.03) \times r_{200}$, in agreement with several previous studies
(e.g. \citep{Newman2013a,Newman2013b}). 

Stellar mass dominates the total mass at $r\lesssim 5-10$\,kpc, while
the cluster outskirts are dominated by DM.  
Such segregation reveals a ``tight coordination'' between the inner DM and
stellar distributions, as also implied by interplay of DM and
baryons that generates the NFW-like total density profile.
The correlation
between the mass inside 5\,kpc and that inside 100\,kpc
(Fig.~\ref{fig:6}) further supports such a tight coordination and points
to similar formation time-scales of the BCG and the inner cluster
region. Thus, the cluster's final configuration depends on the baryonic
content and their formation process \citep{DelPopolo2012a}.
Therefore, in the context of hierarchical structure formation models,
we should expect tight correlations between the final inner baryonic
content and the BCG mass,
as well as between the total baryonic and cluster masses
\citep[see][]{Whiley2008}.

Since the DM and baryon contents sum to the total mass and the baryons  
dominate the inner $5-10$\,kpc region, the inner slope of the DM density
must be shallower than that of the total density,  that is,
$\alpha<1$, as shown in Fig.~\ref{fig:1}.
The observed inner DM slopes $\alpha$ span the range $[0.30, 0.79]$. 

Correlations were also examined between several of the characteristic
quantities of clusters (e.g., $R_\mathrm{e}$, $M_\mathrm{e}$, $M_{500}$).
Our findings are summarized as follows: 
\renewcommand{\theenumi}{\alph{enumi}}
\begin{enumerate}
\item The inner 3D slope of the DM density, $\alpha$, is anti-correlated
      with the BCG effective radius $R_\mathrm{e}$. 
      The anti-correlation reflects the balance between DM and the BCG
      in the cluster center.  
      For an NFW-like total mass profile, clusters with more
      massive BCGs contain less central DM, implying a flatter DM slope. 
\item Similarly, the inner DM slope $\alpha$ and the BCG mass
      $M_\mathrm{e}$ are anti-correlated with each other.
      This indicates again that a larger content of the central baryons
      gives rise to flatter DM profiles.
\item The cluster halo mass $M_{500}$ and the BCG effective radius
      $R_\mathrm{e}$ are correlated with each other, as found in
      previous studies \citep{Kravtsov2013}. 
\item The cluster mass inside 5\,kpc, dominated by the
      stellar baryons, and the cluster mass inside 100\,kpc, dominated
      by DM, are correlated with each other. This hints at early
      formation of the BCG and the inner cluster region, while
      subsequent, continuous mass accretion played a fundamental role in
      the growth of cluster outskirts \citep[e.g.,][]{Fujita2018a,Fujita2018b}.
\end{enumerate}
 
These observed correlations are in support of the physical picture
proposed by \citep{Newman2013a,Newman2013b},
that clusters form from a dissipative
phase that leads to steepening the central stellar density,
followed by a second dissipationless phase in which
interactions between baryonic clumps and DM through dynamical
friction kinematically heat the latter, leading to
flat DM density profiles
\citep{ElZant2001,ElZant2004,RomanoDiaz2008,DelPopolo2012a,DelPopolo2012b,Cole2011,Nipoti2015}.

\section*{Acknowledgments}
To our colleague Lee Xiguo, in memoriam.\\A.D.P. was supported by the Chinese Academy of Sciences and by the
President's International Fellowship Initiative, grant no. 2017
VMA0044.
%K.U. acknowledges support from the Ministry
%of Science and Technology of Taiwan (grant
%MOST 106-2628-M-001-003-MY3) and from the Academia Sinica %Investigator
%Award. 
M.LeD. acknowledges the financial support by Lanzhou University
starting fund. A.D.P. and M.LeD. thank Keiichi Umetsu, who contributed to calculations and improving
the paper, and the referee Stephano Ettori for discussions which
helped improving our paper.

\appendix

\section{Appendix: the model}\label{sec:model}

In this paper, we used, as in several previous studies \citep{DelPopolo2010,DelPopolo2011,DelPopolo2012a,DelPopolo2012b,DelPopolo2014},
a semi-analytical model (SAM) introduced in \citep{DelPopolo2009} and
extended in \citep{DelPopolo2016a,DelPopolo2016b}. 
This model incorporates a secondary infall model (SIM)
\citep{Gunn1972,Hoffman1985,Ascasibar2004,Williams2004} that takes into
account the effect of DM adiabatic contraction
\citep{Blumenthal1986,Gnedin2004,Gustafsson2006}  
of ordered and random angular momentum \citep{Ryden1988,Williams2004},
as well as of baryon-DM angular momentum transfer through dynamical friction
\citep{ElZant2004,Nipoti2004,DelPopolo2009,DelPopolo2009a,DelPopolo2010,Cole2011,Nipoti2015},
in contrast to previous SIMs. 
It also accounts for cooling, reionization, star formation, and SN/AGN
feedback (see the following). 

It starts from the Hubble flow expansion of a perturbation starting in
the linear phase, following its evolution until the maximum expansion
(turn-around) and recollapse to a final density, given by
%{\bf
\citep{Hiotelis2002}%}
%\citep{Gunn1977,Fillmore1984} 
\begin{equation}
\rho(x)=\frac{\rho_\mathrm{ta}(x_\mathrm{m})}{f(x_\mathrm{i})^3}
 \left[1+\frac{d \ln f(x_\mathrm{i})}{d \ln x_\mathrm{m}(x_\mathrm{i})}
 \right]^{-1} 
\label{eq:dturnnn} 
\end{equation}
with $f(x_\mathrm{i})=x/x_\mathrm{i}(x_\mathrm{i})$, the so-called
collapse factor \citep[see Eq. A18,][]{DelPopolo2009}, and 
\begin{equation}
x_\mathrm{m}=g(x_\mathrm{i})=x_{\rm i}\frac{1+\overline{\delta}_\mathrm{i}}
 {\overline{\delta}_\mathrm{i}-(\Omega_\mathrm{i}^{-1}-1)}\;.
\end{equation}
Here $x_\mathrm{m}$ gives the turn-around radius
$x_\mathrm{m}(x_\mathrm{i})$, where the initial density parameter is
$\Omega_\mathrm{i}$, and a given shell's average overdensity reads
$\overline{\delta}_\mathrm{i}$.  

%{\bf 
The collapse factor $f(x_i)$ of a shell with initial radius $x_i$ and apapsis (turn-around) $x_m(x_i)$ can be also written as:
\begin{equation}
f(x_i)=\frac{m_p(x_m)}{m_p(x_m)+m_{add}(x_m)}
\label{eq:cfact}
\end{equation}
where $m_p$ (permanent component) is the mass contained in shells with apapsis smaller than $x_m$, and $m_{add}$ (additional mass) the contribution of the outer shells passing momentarily through the shell $x_m$.
%
%If a shell has an apapsis radius (i.e., apocenter) $x_m$ and initial radius $x_i$, then the mass inside $x_m%$ is obtained summing the mass contained in shells with apapsis smaller than $x_m$ (permanent component, 
%$m_p$) and the contribution of the outer shells passing momentarily through the shell $x_m$ (additional mass %$m_{add}$). 
The total mass is thus given by:
\begin{equation}
m_T(x_m)=m_p(x_m)+m_{add}(x_m)=m_p(x_m)+\int_{x_m}^{R} P_{r_m} (x) \frac{d m(x)}{dx} dx
\label{eq:mpp}
\end{equation}
where $R$ is the radius of the system (the apapsis of the outer shell) and the distribution of mass $m(x)=m(x_m)$,
% is given by Eq. (\ref{eq:dturn}).
while $P_{x_m}(x)$ is the probability to find the shell with apapsis $x$ inside radius $x_m$,
calculated as the ratio of the time the outer shell (with apapsis
$x$) spends inside radius $x_m$ to its period. 
This last quantity can be expressed as 
\begin{equation}
P_{x_m}(x)= \frac{
\int_{x_p}^{x_m} \frac{d \eta}{v_x(\eta)} }
{\int_{x_p}^{x} \frac{d \eta}{v_x(\eta)}
}
\end{equation}
where $x_p$ is the pericenter of the shell with apsis $x$ and $v_x (\eta)$ is the radial velocity of the shell with apapsis $x$ as it passes radius $\eta$.
This radial velocity can be obtained by integrating the equation of motion of the shell:
\begin{equation}
\frac{d v^2}{d t}+2 \mu v^2=2 \left[
\frac{h^2+j^2}{r^3} -G\frac{m_T}{r^2} + \frac{\Lambda}{3} r
\right] v
\label{eq:veloc}
\end{equation}
which can be solved numerically for $v$.

The final density (Eq.~\ref{eq:dturnnn}) is then a complex aggregate depending on several quantities like angular momentum, dynamical friction, etc. 

Summarizing, in order to obtain the density profile for a given shell, knowing initial conditions, angular momentum and coefficient of dynamical friction, one needs to integrate the equation of motions (Eq. \ref{eq:veloc}), then to calculate the probability $P_{x_i}$, 
%(Eq. \ref{eqb11}), 
yielding the contribution of the shell to $m_{add}$ (see Eq. \ref{eq:mpp}), the collapse factor (Eq. \ref{eq:cfact}) and finally the density profile through Eq. (\ref{eq:dturnnn}) (see also \cite[Sect. 4]{Lokas2000}; \cite[Sect. 2.1]{Ascasibar2004}).
%one has to calculate the collapse factor $f$ and the........
%}

The perturbation contains DM and baryons, the latter initially in the gas phase, with the amount set by the cosmic baryon fraction $f_\mathrm{b}=0.17 \pm 0.01$ \citep{Komatsu2009} \citep[while it is 0.167
in][]{Komatsu2011}.
%, itself obtained from the star-formation processes described in the following. 

The tidal torques of larger scales on smaller-scale structures produce the ``ordered'' angular momentum $h$ through the tidal torque theory \citep{Hoyle1953,Peebles1969,White1984,Ryden1988,Eisenstein1995}, while  
random velocities generate ``random'' angular momentum
\citep{Ryden1987}, $j$, expressed in terms of the eccentricity
$e=r_\mathrm{min}/r_\mathrm{max}$
\citep{AvilaReese1998}.
 Here $r_\mathrm{max}$ denotes the apocentric radius, and 
$r_\mathrm{min}$ the pericentric radius.  

The dynamical state of the system induces eccentricity to be corrected
as in simulations of \cite{Ascasibar2004},
\begin{equation}
e(r_\mathrm{max})\simeq 0.8\left(\frac{r_\mathrm{max}}{r_\mathrm{ta}}\right)^{0.1}
\end{equation}
for $r_\mathrm{max}<0.1 r_\mathrm{ta}$.

Dynamical friction was introduced in the equation of motion with a
deceleration term \citep[Eq. A14,][]{DelPopolo2009}, whose coefficient
proceeds as in
\citep{AntonuccioDelogu1994} and
\citep[Appendix D]{DelPopolo2009}. 

Baryon collapse induces adiabatic contraction (AC) of DM, according to
the following mechanism. 
From a protostructure of $f_\mathrm{b}=M_\mathrm{b}/M_{500} \ll 1$
baryons and $1-f_\mathrm{b}$\footnote{The gas mass, $M_\mathrm{gas}$, and mass
in stars, $M_\mathrm{*}$, combine into the total baryonic mass,
$M_\mathrm{b}$.} DM, the baryons cool and collapse toward the halo center,
forming the distribution $M_\mathrm{b}( r)$.
This compresses DM, relocating particles from $r_\mathrm{i}$ to 
\begin{equation}
r \left [ M_\mathrm{b}(r) +M_\mathrm{DM}(r) \right] = r_\mathrm{i} M_\mathrm{i} (r_\mathrm{i})
\label{eq:ad1}
\end{equation}
\cite{Blumenthal1986},
with $M_\mathrm{i}(r_\mathrm{i})$ the initial total mass
and
$M_\mathrm{DM}$ the final DM distribution.
%%%
Assuming equal initial baryon and DM distributions
\citep{Mo1998,Cardone2005,Treu2002,Keeton2001} and the final Hernquist
baryon distribution \citep{Rix1997,Keeton2001,Treu2002},
$M_\mathrm{i}(r_\mathrm{i})$ and
$M_\mathrm{b}(r)$ are given,
and the DM mass distribution in the absence of
shell crossing is obtained by solving Eqs. (\ref{eq:ad1}) and
(\ref{eq:ad2}), 
\begin{equation}
M_ \mathrm{DM} (r)=(1-F_\mathrm{b}) M_\mathrm{i} (r_\mathrm{i})
\label{eq:ad2}
\end{equation}
to find the final halo distribution.
Conservation of angular momentum,
represented by the product of the orbit-averaged radius $\bar{r}$ with
its inner mass \citep{Gnedin2004}, 
\begin{equation}
  M(\bar{r})\bar{r}= \mathrm{const.},
  \label{eq:modified}
\end{equation}
using 
\begin{equation}
  \bar{r} = {2 \over T_r} \int_{r_\mathrm{min}}^{r_\mathrm{max}} r \, {dr \over v_r},
\end{equation}
and $T_r$ as the radial period, improves the model.

Our treatment of star formation, gas cooling, reionization, and
supernovae feedback (SNF) follows
\cite[Secs.~2.2.2 and 2.2.3]{DeLucia2008,Li2010}
  
With reionization as in \cite{Li2010}, the baryon fraction is reduced in 
the redshift range $z=[11.5, 15]$ and modified as
\begin{equation}
 f_\mathrm{b, halo}(z,M_{\rm Vir})=\frac{f_\mathrm{b}}{[1+0.26 M_\mathrm{F}(z)/M_{\rm Vir}]^3}\;,
\end{equation}
calculated with $M_\mathrm{F}$ the ``filtering'' mass
\citep*[see][]{Kravtsov2004} and the virial mass, $M_{\rm Vir}$ ($\Delta=200$) was converted to the cluster halo mass  $M_{500}$ following \citep{Hu2003}.
%({\bf KU: I assumed that what you mean by Mvir is M500}). 
Although the \cite{Ryden1988} treatment yields similar results, a
classical cooling flow \citep[e.g.,][]{White1991} \citep[see Sec.~2.2.2
of][]{Li2010} is used here for gas cooling. 

We follow \cite{DeLucia2008} to account for star-formation
processes. SNF follows the treatment of
\cite{Croton2006}. The blast-wave SNF \citep{Stinson2006} was used in
\cite{DiCintio2014}. Although our choice of formalism is similar, it is
not very fundamental, and our model differs from their SNF model
\citep[e.g.,][]{DiCintio2014} in the occurrence of the flattening
process before star formation and in our gravitational source of
energy, whereas the SNF flattening process occurs after star formation 
with stellar feedback, which is in act as the energy source after core
formation disrupting the gas clouds.

The stellar formation occurs once a disc is formed from gas with the star formation rate
\begin{equation}
\psi=0.03 M_\mathrm{sf}/t_\mathrm{dyn},
\end{equation}
which is computed with the disc dynamical time, $t_\mathrm{dyn}$, and the
gas mass $M_\mathrm{sf}$ contained in regions where its density is above a
given density threshold, $n>9.3$\,cm$^{-3}$, in the same way as in
\cite{DiCintio2014}. We use the Chabrier \citep{Chabrier2003} initial
mass function (IMF), forming stars with
\begin{equation}
 \Delta M_{\ast}=\psi\Delta t
\end{equation}
per units of time $\Delta t$.

SNF injects energy in the interstellar medium (ISM) at a rate of
\begin{equation}
 \Delta E_\mathrm{SN}=0.5\epsilon_\mathrm{halo}\,\Delta M_{\ast} \,V^2_\mathrm{SN},
\end{equation}
obtained with the energy injected per supernova and per unit solar mass,
$V^2_\mathrm{SN}=\eta_\mathrm{SN}E_\mathrm{SN}$. 
The fixed efficiency $\epsilon_{\rm halo}=0.35$ \citep{Li2010} of the
disc gas reheating by this energy injection is computed with the
supernova number per solar mass, $\eta_\mathrm{SN}=8\times
10^{-3}/M_{\odot}$, assuming a Chabrier IMF \citep{Chabrier2003}, and
with the typical energy released in a SN explosion,
$E_\mathrm{SN}=10^{51}$\,erg. 

This energy injection reheats the gas in proportion to the star-formation number,
\begin{equation}
 \Delta M_{\rm reheat} = 3.5 \Delta M_{\ast},
\end{equation}
inducing a thermal energy change 
%({\bf KU: What is $V^2_\mathrm{vir}
%\mathrm{vir}$
%(undefined)? Or do you mean $V^2_\mathrm{SN}?$})
\begin{equation}
 \Delta E_\mathrm{hot}= 0.5\Delta M_\mathrm{reheat} V^2_{Vir},
\end{equation}
where $V_{Vir}= (10 G H(z) M_{Vir})^{1/3}$ with $H(z)$ being the Hubble
constant at redshift $z$.  

The last will eject a quantity of the hot gas equal to 
\begin{equation}
 \Delta M_\mathrm{eject}=\frac{\Delta E_\mathrm{SN}-\Delta E_\mathrm{hot}}{0.5 V^2_{Vir}}
\end{equation}
from the halo with the condition $\Delta E_\mathrm{SN}>\Delta E_\mathrm{hot}$.

All the quantitities evaluated at virial radius (e.g., $M_{\rm Vir}, V_{\rm Vir}$) were converted to corresponding quantities at $\Delta=500$, by following \citep{Hu2003}.

Subsequently, the hot gas can be accreted by the halo into its hot
component in link with the central galaxy
\citep{DeLucia2004,Croton2006}. 

Accounting for AGN quenching is required for masses $M\simeq 6 \times
10^{11} M_{\odot}$  \citep{Cattaneo2006}. Here, we use the prescription of
\cite{Martizzi2012,Martizzi2012a} to implement AGN feedback: A
conjunction of the star density above  $2.4 \times 10^6 M_{\odot}$\,kpc$^{-3}$,
the gas density at 10 times the stellar density,
and the 3D velocity dispersion exceeding 100\,km\,s$^{-1}$
creates an initial (seed) super-massive black hole (SMBH) with
$10^5M_{\odot}$, which grows with accretion and produces AGN feedback
following a variant of the \cite{Booth2009} model 
%({\bf KU:Reference for Booth2009 missing in the bib file})
modified according to \cite{Martizzi2012}. 

%{\bf
  Our model is a semi-analytic simulation (SAM), such as, e.g., GALFORM, and GALACTICUS. 
%There is not a single parameter, M500. 
Its parameters are the same found in (e.g., SPH) simulations (e.g., baryon fraction fixed as the cosmic baryon fraction value, a parameter fixing the random angular momentum, set as in simulations, star formation rate, density treshold of star formation, parameters related to supernovae and AGN feedback and other parameters discussed in the model section). 
A more complete description of the model can be found in \cite{DelPopolo2016b}
%Del Popolo et al. (2018) (PRD 98, 063517)
, and \cite{DelPopolo2018}.
%Del Popolo (2016) (Astrophys Space Sci 361:222). 
%}

Finally, the robustness of the model has been demonstrated in various manners as
summarized below.
\renewcommand{\theenumi}{\alph{enumi}}
\begin{enumerate}
 \item  DM heating by collapsing baryonic clumps that induce cusp
	flattening was predicted in 2009 for galaxies, and 2012 for
	clusters.
	The predictions are in agreement with the studies of
	\citep{ElZant2001,ElZant2004,RomanoDiaz2008,RomanoDiaz2009,Cole2011,Inoue2011,Nipoti2015}. Our
	model is compared with the smoothed particle hydrodynamics (SPH)
	simulations of \cite{Governato2010} in Fig.~4 of \citep{DelPopolo2011}.
 \item Correct predictions for the galaxy density profiles
       \citep{DelPopolo2009,DelPopolo2009a} and the cluster density
       profiles \citep{DelPopolo2012b} were made before SPH simulations
       of \cite{Governato2010,Governato2012} 
       and  \cite{Martizzi2013}, respectively. 
\item The dependence of the inner DM density slope on the halo mass
      \cite[Fig.~2a, solid 
      line]{DelPopolo2010} was predicted before the quasi similar
      results of \cite[][Fig.~6, solid lines]{DiCintio2014} presented in
      terms of the rotation velocity, 
      $V_\mathrm{c}$ 
      \citep[$V_\mathrm{c}=2.8 \times 10^{-2} M_\mathrm{vir}^{0.316}$,][]{Klypin2011}.
\item The dependence of the cluster baryon fraction on the inner
      DM slope was also predicted in \cite{DelPopolo2012b}, which
      was later studied and confirmed
%      ({\bf KU: and confirmed??}) 
in SPH simulations \cite{DiCintio2014}.
\item  Figure 1 of
      \citep{DelPopolo2016a,DelPopolo2016b} compares the dependence of
       the inner DM slope on the halo mass with the
       \citep{DiCintio2014} simulations.
      Predictions for the Tully--Fisher relation, the Faber--Jackson relation, and the
       $M_{*}-M_\mathrm{halo}$, relation are compared with simulations
       in Figs.~4 and 5 of \citep{DelPopolo2016a,DelPopolo2016b}. 
\end{enumerate}

%The model demonstrated its robustness predicting before simulations the density flattening and shape %produced by heating of DM, for galaxies \cite{DelPopolo2009,DelPopolo2010,DelPopolo2012a} in agreement with %{subsequent} SPH simulations \citep[e.g.][]{Governato2010,Governato2012} 
%\cite[see also Fig.~4 of][for a direct comparison]{DelPopolo2011}, and similarly for clusters of galaxies  
%\cite{DelPopolo2012b} in agreement with hydro-simulations of \citep{Martizzi2013}, the inner density slope %dependence on the halo mass and on the total baryonic content to the total mass ratio 
%\cite{DelPopolo2010,DelPopolo2016a}, in 
%agreement with \cite{DiCintio2014}. The slope was shown by \cite{DelPopolo2010,DelPopolo2016a} to depend %also on the angular momentum. In \cite{DelPopolo2016b}, the stellar and baryonic Tully-Fisher, Faber-Jackson %and the stellar mass vs. halo mass (SMH) relations were shown in agreement with simulations.
%
%Finally, the correct DM profile inner slope dependence on halo mass was explained over 6 order of magnitudes %in halo mass, from dwarves to clusters\cite{DelPopolo2009,DelPopolo2010,DelPopolo2012a,DelPopolo2012b}, a %range no other model achieved.
%}

\bibliographystyle{elsarticle-num}

\bibliography{MasterBib}

\end{document}